\documentclass[pre,twocolumn,aps,showpacs]{revtex4}
\newcommand{\bec}[1]{\mbox{\boldmath $ #1$}}
\usepackage{graphicx}
\begin{document}
\title{Formation of large-scale semi-organized structures in
turbulent convection}
\author{Tov Elperin}
\email{elperin@menix.bgu.ac.il}
\homepage{http://www.bgu.ac.il/~elperin}
\author{Nathan Kleeorin}
\email{nat@menix.bgu.ac.il}
\author{Igor Rogachevskii}
\email{gary@menix.bgu.ac.il}
\homepage{http://www.bgu.ac.il/~gary}
\affiliation{Department of Mechanical Engineering, The Ben-Gurion
University of the Negev, \\
POB 653, Beer-Sheva 84105, Israel}
\author{Sergej Zilitinkevich}
\email{sergej@met.uu.se}
\affiliation{Department of Earth
Sciences, Meteorology,
Uppsala University, \\
Villavagen 16, S-752 36 Uppsala, Sweden}
\date{\today}
\begin{abstract}
A new mean-field theory of turbulent convection is developed by
considering only the small-scale part of spectra as "turbulence"
and the large-scale part, as a "mean flow", which includes both
regular and semi-organized motions. The developed theory predicts
the convective wind instability in a shear-free turbulent
convection. This instability causes formation of large-scale
semi-organized fluid motions in the form of cells or rolls.
Spatial characteristics of these motions, such as the minimum size
of the growing perturbations and the size of perturbations with
the maximum growth rate, are determined. This study predicts also
the existence of the convective shear instability in a sheared
turbulent convection. This instability causes generation of
convective shear waves which have a nonzero hydrodynamic helicity.
Increase of shear promotes excitation of the convective shear
instability. Applications of the obtained results to the
atmospheric turbulent convection and the laboratory experiments on
turbulent convection are discussed. This theory can be applied
also for the describing a mesogranular turbulent convection in
astrophysics.
\end{abstract}

\pacs{47.65.+a; 47.27.-i}

\maketitle

\section{Introduction}

In the last decades it has been recognized that the very high
Rayleigh number convective boundary layer (CBL) has more complex
nature than might be reckoned. Besides the fully organized
component naturally considered as the mean flow and the chaotic
small-scale turbulent fluctuations, one more type of motion has
been discovered, namely, long-lived large-scale structures, which
are neither turbulent nor deterministic (see, e.g.,
\cite{EB93,AZ96,LS80,H84,W87,H88,SS89,SH89,M91,Z91,WH92,WK96,ZGH98,YKH02}).
These semi-organized structures considerably enhance the vertical
transports and render them essentially non-local in nature. In the
atmospheric shear-free convection, the structures represent
three-dimensional Benard-type cells composed of narrow uprising
plumes and wide downdraughts. They embrace the entire convective
boundary layer ($\sim 2$ km in height) and include pronounced
large-scale ($\sim 5$ km in diameter) convergence flow patterns
close to the surface (see, e.g., \cite{EB93,AZ96}, and references
therein). In sheared convection, the structures represent
CBL-scale rolls stretched along the mean wind. Life-times of the
semi-organized structures are much larger than the turbulent time
scales. Thus, these structures can be treated as comparatively
stable, quasi-stationary motions, playing the same role with
respect to small-scale turbulence as the mean flow.

In a laboratory turbulent convection several organized features of
motion, such as plumes, jets, and the large-scale circulation, are
known to exist. The experimentally observed large-scale
circulation in the closed box with a heated bottom wall (the
Rayleigh-Benard apparatus) is often called the "mean wind" (see,
e.g., \cite{KH81,SWL89,CCL96,AS99,K01,NSS01,S94,BBC86}, and
references therein). There are several unsolved theoretical
questions concerning these flows, e.g., how do they arise, and
what are their characteristics and dynamics.

In spite of a number of studies, the nature of large-scale
semi-organized structures is poorly understood. The Rayleigh
numbers, $ {\rm Ra} ,$ based on the molecular transport
coefficients are very large (of the order of $ 10^{11} - 10^{13}
).$ This corresponds to fully developed turbulent convection in
atmospheric and laboratory flows. At the same time the effective
Rayleigh numbers, $ {\rm Ra}^{\rm (eff)} ,$ based on the turbulent
transport coefficients (the turbulent viscosity and turbulent
diffusivity) are not high, e.g., $ {\rm Ra}^{\rm (eff)} \sim {\rm
Ra} / ({\rm Re} {\rm Pe}) ,$ where $ {\rm Re} $ and $ {\rm Pe} $
are the Reynolds and Peclet numbers, respectively. They are less
than the critical Rayleigh numbers required for the excitation of
large-scale convection. Hence the emergence of large-scale
convective flows (which are observed in the atmospheric and
laboratory flows) seems puzzling.

The main goal of this study is to suggest a mechanism for
excitation of large-scale circulations (large-scale convection).
In particular, in the present paper we develop a new mean-field
theory of turbulent convection by considering only the small-scale
part of spectra as "turbulence" and the large-scale part, as a
"mean flow", which includes both, regular and semi-organized
motions. We found a convective wind instability in a shear-free
turbulent convection which results in formation of large-scale
semi-organized fluid motions in the form of cells or rolls
(convective wind). We determined the spatial characteristics of
these motions, such as the minimum size of the growing
perturbations and the size of perturbations with the maximum
growth rate. In addition, we studied a convective shear
instability in a sheared turbulent convection which causes a
generation of convective shear waves. We analyzed the relevance of
the obtained results to the turbulent convection in the atmosphere
and the laboratory experiments.

Traditional theoretical models of the boundary-layer turbulence,
such as the Kolmogorov-type  closures and similarity theories
(e.g., the Monin-Obukhov surface-layer similarity theory) imply
two assumptions: (i) Turbulent flows can be decomposed into two
components of principally different nature: fully organized
(mean-flow) and fully turbulent flows. (ii) Turbulent fluxes are
uniquely determined by the local mean gradients. For example, the
turbulent flux of entropy is given by
\begin{eqnarray}
\langle s {\bf u} \rangle = - \kappa_T \bec{\nabla} \bar S \;
\label{AA1}
\end{eqnarray}
(see, e.g., \cite{MY75}), where $\kappa_T$  is the turbulent
thermal conductivity, $\bar S$ is the mean entropy, ${\bf u}$ and
$s$ are fluctuations of the velocity and entropy.

However, the mean velocity gradients can affect the turbulent flux
of entropy. The reason is that additional essentially
non-isotropic velocity fluctuations can be generated by tangling
of the mean-velocity gradients with the Kolmogorov-type
turbulence. The source of energy of this "tangling turbulence" is
the energy of the Kolmogorov turbulence.

In the present paper we showed that the tangling turbulence can
cause formation of semi-organized structures due to excitation of
large-scale instability. The tangling turbulence was introduced by
Wheelon \cite{W57} and Batchelor et al. \cite{BH59} for a passive
scalar and by Golitsyn \cite{G60} and Moffatt \cite{M61} for a
passive vector (magnetic field). Anisotropic fluctuations of a
passive scalar (e.g., the number density of particles or
temperature) are generated by tangling of gradients of the mean
passive scalar field with random velocity field. Similarly,
anisotropic magnetic fluctuations are excited by tangling of the
mean magnetic field with the velocity fluctuations. The Reynolds
stresses in a turbulent flow with a mean velocity shear is another
example of a tangling turbulence. Indeed, they are strongly
anisotropic in the presence of shear and have a steeper spectrum $
(\propto k^{-7/3}) $ than a Kolmogorov turbulence (see, e.g.,
\cite{L67,WC72,SV94,IY02}). The anisotropic velocity fluctuations
of tangling turbulence were studied first by Lumley \cite{L67}.

This paper is organized as follows. In Section II we described the
governing equations and the method of the derivations of the
turbulent flux of entropy and Reynolds stresses. In Section III
using the derived mean field equations we studied the large-scale
instability in a shear-free turbulent convection which causes
formation of semi-organized fluid motions in the form of cells. In
Section IV the instability in a sheared turbulent convection is
investigated and formation of large-scale semi-organized rolls is
described. Application of the obtained results for the analysis of
observed semi-organized structures in the atmospheric turbulent
convection is discussed in Section V.

\section{The governing equations and the method of the derivations}

Our goal is to study the tangling turbulence, in particular, an
effect of sheared large-scale motions on a developed turbulent
stratified convection. To this end we consider a fully developed
turbulent convection in a stratified non-rotating fluid with large
Rayleigh and Reynolds numbers. The governing equations read:
\begin{eqnarray}
\biggl({\partial \over \partial t} + {\bf v} \cdot
\bec{\nabla}\biggr){\bf v} &=& - \bec{\nabla} \biggl({P \over
\rho_0}\biggr) - {\bf g} S + {\bf f}_{\nu}({\bf v}) \;,
\label{B1} \\
\biggl({\partial \over \partial t} + {\bf v} \cdot
\bec{\nabla}\biggr) S &=& - {\bf v} \cdot {\bf N}_{b} - {1 \over
T_{0}} \, \bec{\nabla} \cdot {\bf F}_{\kappa}(S) \;, \label{B2}
\end{eqnarray}
where $ {\bf v} $ is the fluid velocity with $\bec{\nabla} \cdot
{\bf v} = {\bf \Lambda} \cdot {\bf v} ,$ $ \, {\bf g} $ is the
acceleration of gravity, $ \rho_0 {\bf f}_{\nu}({\bf v}) $ is the
viscous force, $ {\bf F}_{\kappa}(S) $ is the heat flux that is
associated with the molecular heat conductivity $ \kappa ,$ $ \,
{\bf \Lambda} = - \rho_0^{-1} \bec{\nabla} \rho_0 $  is the
density stratification scale, and $ {\bf N}_{b} = (\gamma
P_{0})^{-1} \bec{\nabla} P_{0} - \rho_0^{-1} \bec{\nabla} \rho_0
.$ The variables with the subscript $ "0" $ correspond to the
hydrostatic equilibrium $ \bec{\nabla} P_{0} = \rho_0 {\bf g} ,$
and $ T_{0} $ is the equilibrium fluid temperature, $ S = P /
\gamma P_{0} -  \rho / \rho_0 $ are the deviations of the entropy
from the hydrostatic equilibrium value, $ P $ and $ \rho $ are the
deviations of the fluid pressure and density from the hydrostatic
equilibrium. Note that the variable $ S = \Theta / \Theta_0 ,$
where $ \Theta $ is the potential temperature which is used in
atmospheric physics. The Brunt-V\"{a}is\"{a}l\"{a} frequency, $
\Omega_{b} ,$ is determined by the equation $ \Omega_{b}^{2} = -
{\bf g} \cdot {\bf N}_{b} .$ In order to derive Eq.~(\ref{B1}) we
used an identity: $ - \bec{\nabla} P + {\bf g} \rho = - \rho_0
[\bec{\nabla} (P / \rho_0) + {\bf g} S - P {\bf N}_{b}/ \rho_0] ,$
where we assumed that $ |P {\bf N}_{b}/ \rho_0| \ll |{\bf g} S| $
and $ |P {\bf N}_{b}/ \rho_0| \ll |\bec{\nabla} (P / \rho_0)| .$
This assumption corresponds to a nearly isentropic basic reference
state when $ {\bf N}_{b} $ is very small. For the derivation of
this identity we also used the equation for the hydrostatic
equilibrium. Equations (\ref{B1}) and (\ref{B2}) are written in
the Boussinesq approximation for $ \bec{\nabla} \cdot {\bf v}
\not= 0 .$

\subsection{Mean field approach}

We use a mean field approach whereby the velocity, pressure and
entropy are separated into the mean and fluctuating parts: $ {\bf
v} = \bar{\bf U} + {\bf u} ,$ $ \, P = \bar P + p ,$ and $ S =
\bar S + s ,$ the fluctuating parts have zero mean values, $
\bar{\bf U} = \langle {\bf v} \rangle ,$ $ \, \bar P = \langle P
\rangle $ and  $ \bar S = \langle S \rangle .$ Averaging
Eqs.~(\ref{B1}) and (\ref{B2}) over an ensemble of fluctuations we
obtain the mean-field equations:
\begin{eqnarray}
\biggl({\partial  \over \partial t} + \bar {\bf U} \cdot
\bec{\nabla}\biggr) \bar U_{i} &=& - {\nabla}_{i} \biggl({\bar P
\over \rho_0}\biggr) + (\Lambda_{j} - {\nabla}_{j}) \langle u_{i}
u_{j} \rangle
\nonumber \\
& & - {\bf g} \bar S + \bar {\bf f}_{\nu}(\bar {\bf U}) \;,
\label{B3} \\
\biggl({\partial  \over \partial t} + \bar {\bf U} \cdot
\bec{\nabla} \biggr) \bar S &=& - \bar {\bf U} \cdot {\bf N}_{b} +
(\Lambda_{i} - {\nabla}_{i}) \langle s \, u_{i} \rangle
\nonumber \\
& & - {1 \over T_{0}} \, \bec{\nabla} \cdot \bar {\bf
F}_{\kappa}(\bar {\bf U},\bar S) \;, \label{B4}
\end{eqnarray}
where $ \rho_0 \bar{\bf f}_{\nu}(\bar {\bf U}) $ is the mean
molecular viscous force, $ \bar{\bf F}_{\kappa}(\bar {\bf U},\bar
S) $ is the mean heat flux that is associated with the molecular
thermal conductivity. In order to derive a closed system of the
mean-field equations we have to determine the mean-field
dependencies of the Reynolds stresses $ f_{ij}(\bar {\bf U},\bar
S) = \langle u_{i}(t,{\bf x}) u_{j}(t,{\bf x}) \rangle $ and the
flux of entropy $ \Phi_{i}(\bar {\bf U},\bar S) = \langle s(t,{\bf
x}) u_{i}(t,{\bf x}) \rangle .$ To this end we used equations for
the fluctuations $ {\bf u}(t,{\bf r}) $ and $ s(t,{\bf r}) $ which
are obtained by subtracting equations (\ref{B3}) and (\ref{B4})
for the mean fields from the corresponding equations (\ref{B1})
and (\ref{B2}) for the total fields:
\begin{eqnarray}
{\partial {\bf u} \over \partial t} &=& - (\bar{\bf U} \cdot
\bec{\nabla}) {\bf u} - ({\bf u} \cdot \bec{\nabla}) \bar{\bf U} -
\bec{\nabla} \biggl({p \over \rho_{0}}\biggr)
\nonumber\\
& & - {\bf g} \, s + {\bf U}_{N} \;,
\label{B5} \\
{\partial s \over \partial t} &=& - {\bf u} \cdot ({\bf N}_{b} +
\bec{\nabla}\bar S) - (\bar{\bf U} \cdot \bec{\nabla}) s + S_{N}
\;, \label{B6}
\end{eqnarray}
where $ U_{N} = \langle ({\bf u} \cdot \bec{\nabla}) {\bf u}
\rangle - ({\bf u} \cdot \bec{\nabla}) {\bf u} + {\bf
f}_{\nu}({\bf u}) $ and $ S_{N} = \langle ({\bf u} \cdot
\bec{\nabla}) s \rangle - ({\bf u} \cdot \bec{\nabla}) s - (1/
T_{0}) \, \bec{\nabla} \cdot {\bf F}_{\kappa}({\bf u}, s) $ are
the nonlinear terms which include the molecular dissipative terms.

\subsection{Method of derivations}

By means of Eqs.~(\ref{B5}) and (\ref{B6}) we determined the
dependencies of the second moments $ f_{ij}(\bar {\bf U},\bar S)$
and $ \Phi_{i}(\bar {\bf U},\bar S) $ on the mean-fields $ \bar
{\bf U} $ and $ \bar S .$ The procedure of the derivation is
outlined in the following (for details see Appendix A).

(a). Using Eqs.~(\ref{B5}) and (\ref{B6}) we derived equations for
the following second moments:
\begin{eqnarray}
f_{ij}({\bf k}) &=& \hat L(u_{i},u_{j}) \;, \quad \Phi_{i}({\bf
k}) = \hat L(s,u_{i}) \;,
\label{B60} \\
F({\bf k}) &=& \hat L(s,\omega) \;, \quad G({\bf k}) = \hat
L(\omega,\omega) \;,
\label{TB61}\\
H({\bf k}) &=& \hat L(s,s) \;, \label{B61}
\end{eqnarray}
where $ \hat L(a,b) = \langle a({\bf k}) b(-{\bf k}) \rangle ,$ $
\,\omega = (\bec{\nabla} {\bf \times} {\bf u})_z ,$ the
acceleration of gravity $ {\bf g} $ is directed opposite to the
$z$ axis. Here we used a two-scale approach. This implies that we
assumed that there exists a separation of scales, i.e., the
maximum scale of turbulent motions $ l_0 $ is much smaller then
the characteristic scale of inhomogeneities of the mean fields.
Our final results showed that this assumption is indeed valid. The
equations for the second moments (\ref{B60})-(\ref{B61}) are given
by Eqs.~(\ref{A6}), (\ref{A7}) and (\ref{A10})-(\ref{A12}) in
Appendix A. In the derivation we assumed that the inverse density
stratification scale $ \Lambda^{2} \ll k^{2} .$

(b). The derived equations for the second moments contain the
third moments, and a problem of closing the equations for the
higher moments arises. Various approximate methods have been
proposed for the solution of problems of this type (see, {\em
e.g.,} \cite{MY75,O70,Mc90}). The simplest procedure is the $ \tau
$ approximation which was widely used for study of different
problems of turbulent transport (see, e.g.,
\cite{O70,PFL76,KRR90,RK2000}). One of the simplest procedures
that allows us to express the third moments $ f^{N}_{ij} ,$ $ {\bf
\Phi}^{N} , \ldots ,$ $ H_{N} $ in Eqs.~(\ref{A6}), (\ref{A7}) and
(\ref{A12}) in terms of the second moments, reads
\begin{eqnarray}
A_N({\bf k}) - A_N^{(0)}({\bf k}) &=& - {A({\bf k}) - A^{(0)}({\bf
k}) \over \tau (k)} \;,
\label{B7}
\end{eqnarray}
where the superscript $ (0) $ corresponds to the background
turbulent convection ({\em i.e.}, a turbulent convection with $
\nabla_{i} \bar U_{j} = 0) ,$ and $ \tau (k) $ is the
characteristic relaxation time of the statistical moments. Note
that we applied the $ \tau $-approximation (\ref{B7}) only to
study the deviations from the background turbulent convection
which are caused by the spatial derivatives of the mean velocity.
The background turbulent convection is assumed to be known.

The $ \tau $-approximation  is in general similar to Eddy Damped
Quasi Normal Markowian (EDQNM) approximation. However there is a
principle difference between these two approaches (see
\cite{O70,Mc90}). The EDQNM closures do not relax to the
equilibrium, and this procedure does not describe properly the
motions in the equilibrium state. Within the EDQNM theory, there
is no dynamically determined relaxation time, and no slightly
perturbed steady state can be approached \cite{O70}. In the $ \tau
$-approximation, the relaxation time for small departures from
equilibrium is determined by the random motions in the equilibrium
state, but not by the departure from equilibrium \cite{O70}.
Analysis performed in \cite{O70} showed that the $ \tau
$-approximation describes the relaxation to the equilibrium state
(the background turbulent convection) more accurately than the
EDQNM approach.

(c). We assumed that the characteristic times of variation of the
second moments $ f_{ij}({\bf k}) ,$ $ \Phi_{i}({\bf k}) , \ldots
,$ $ H({\bf k}) $ are substantially larger than the correlation
time $ \tau(k) $ for all turbulence scales. This allowed us to
determine a stationary solution for the second moments $
f_{ij}({\bf k}) ,$ $ \Phi_{i}({\bf k}) , \ldots ,$ $ H({\bf k}) .$

(d). For the integration in $ {\bf k} $-space of the second
moments $ f_{ij}({\bf k}) ,$ $ \Phi_{i}({\bf k}) , \ldots ,$ $
H({\bf k}) $ we have to specify a model for the background
turbulent convection. Here we used the following model of the
background turbulent convection which is discussed in more details
in Appendix B:
\begin{eqnarray}
f_{ij}^{(0)}({\bf k}) &=& f_{\ast} [P_{ij}({\bf k}) + \varepsilon
P_{ij}^{(\perp)}({\bf k}_{\perp})] \tilde W(k) \;,
\label{B8} \\
\Phi^{(0)}_{i}({\bf k}) &=& k_{\perp}^{-2} [k^2
\Phi_{z}^{(0)}({\bf k}) e_{j} P_{ij}({\bf k})
\nonumber\\
& &+ i F^{(0)}({\bf k}) ({\bf e} {\bf \times} {\bf k})_{i}] \;,
\label{B9} \\
\Phi_{z}^{(0)}({\bf k}) &=& {\Phi}^{\ast}_z \biggl[2 \alpha - 3
(\alpha - 1)  \biggl({k_{\perp} \over k}\biggr)^{2} \biggr] \tilde
W(k) \;,
\label{B10} \\
F^{(0)}({\bf k}) &=& - 6 i f^{(0)}({\bf k}) ({\bf \Phi}^{\ast}
\cdot ({\bf e} {\bf \times} {\bf k})) \;,
\label{B11} \\
G^{(0)}({\bf k}) &=& (1 + \varepsilon) f_{\ast} f^{(0)}({\bf k})
k^{2} \;,
\label{MB12} \\
H^{(0)}({\bf k}) &=&  2 H_{\ast} \tilde W(k) \;,
\label{B12}
\end{eqnarray}
where $ \tilde W(k) = W(k) / 8 \pi k^{2} ,$ $ \, f^{(0)}({\bf k})
= (k_{\perp} / k)^{2} \tilde W(k) ,$ $ \, \varepsilon = (2/3)
[\langle {\bf u}_{\perp}^{2} \rangle / \langle {\bf u}_{z}^{2}
\rangle - 2] $ is the degree of anisotropy of the turbulent
velocity field $ {\bf u} = {\bf u}_{\perp} + u_{z} {\bf e} ,$ $ \,
\alpha $ is the degree of anisotropy of the turbulent flux of
entropy (see below and Appendix B), $ P_{ij}({\bf k}) =
\delta_{ij} - k_{ij} ,$ $ \, k_{ij} = k_{i} k_{j} / k^{2} ,$ $ \,
{\bf k} = {\bf k}_{\perp} + k_{z} {\bf e} ,$ $ \, k_{z} = {\bf k}
\cdot {\bf e} ,$ $ \, P_{ij}^{(\perp)}({\bf k}_{\perp}) =
\delta_{ij} - k^{\perp}_{ij} - e_{ij} ,$ $ \, k^{\perp}_{ij} =
({\bf k}_{\perp})_{i} ({\bf k}_{\perp})_{j} / k_{\perp}^{2} ,$ $
\, e_{ij} = e_{i} e_{j} , \,$ ${\bf e}$ is the unit vector
directed along the $z$ axis. Here $ \tau(k) = 2 \tau_0 \bar
\tau(k) ,$ $ \, W(k) = - d \bar \tau(k) / dk ,$ $ \, \bar \tau(k)
= (k / k_{0})^{1-q} ,$ $ \, 1 < q < 3 $  is the exponent of the
kinetic energy spectrum $ (q = 5/3 $ for Kolmogorov spectrum), $
k_{0} = 1 / l_{0} ,$ and $ l_{0} $ is the maximum scale of
turbulent motions, $ \tau_0 = l_{0} / u_{0} $ and $ u_{0} $ is the
characteristic turbulent velocity in the scale $ l_{0} .$ Motion
in the background turbulent convection is assumed to be
non-helical. In Eqs.~(\ref{B8}) and~(\ref{B9}) we neglected small
terms $ \sim O(\Lambda f_{\ast}; \bec{\nabla} f_{\ast}) $ and $
\sim O(\Lambda \Phi^{\ast}; \bec{\nabla} \Phi^{\ast}) ,$
respectively. Note that $ f_{ij}^{(0)}({\bf k}) e_{ij} = f_{\ast}
f^{(0)}({\bf k}) .$ Now we calculate $ f_{ij}^{(0)} \equiv \int
f_{ij}^{(0)}({\bf k}) \,d {\bf k} $ using Eq.~(\ref{B8}):
\begin{eqnarray}
f_{ij}^{(0)} = f_{\ast} \biggl[{1 \over 3} \delta_{ij} +
{\varepsilon \over 4} (\delta_{ij} - e_{ij}) \biggr] \; .
\label{B14}
\end{eqnarray}
Note that $ {\bf \Phi}^{(0)} \equiv \int {\bf \Phi}^{(0)}({\bf k})
\,d {\bf k} = {\bf \Phi}^{\ast} .$ The parameter $ \alpha $ can be
presented in the form
\begin{eqnarray}
\alpha &=& {1 + \xi (q + 1) / (q - 1) \over 1 + \xi / 3} \;,
\label{D6} \\
\xi  &=& (l_{\perp} / l_{z})^{q-1} - 1 \;,
\label{D7}
\end{eqnarray}
where $ l_{\perp} $ and $ l_{z} $ are the horizontal and vertical
scales in which the correlation function $ \Phi_{z}^{(0)}({\bf r})
= \langle s({\bf x}) \, {\bf u}({\bf x}+{\bf r}) \rangle $ tends
to zero (see Appendix B). The parameter $ \xi $ describes the
degree of thermal anisotropy. In particular, when $ l_{\perp} =
l_{z} $ the parameter $ \xi = 0 $ and $ \alpha = 1 .$ For $
l_{\perp} \ll l_{z} $ the parameter $ \xi = - 1 $ and $ \alpha = -
3  / (q-1) .$ The maximum value $ \xi_{\rm max} $ of the parameter
$ \xi $ is given by $ \xi_{\rm max} = q - 1 $ for $ \alpha = 3 .$
Thus, for $ \alpha < 1 $ the thermal structures have the form of
column or thermal jets $ (l_{\perp} < l_{z}) ,$ and for $ \alpha >
1 $ there exist the `'pancake'' thermal structures $ (l_{\perp} >
l_{z}) $ in the background turbulent convection. For statistically
stationary small-scale turbulence the degree of anisotropy of
turbulent velocity field varies in the range
\begin{eqnarray}
- {\rm min} \, \biggl\{{4 (q +3) \over 5 (q + 1)}; \, {2 (19 - q)
\over 25} ; \, {4 \over 3} \biggr \} < \varepsilon < \infty \; .
\label{C37}
\end{eqnarray}
The negative (positive) degree of anisotropy $ \varepsilon $ of a
turbulent velocity field corresponds to that the vertical size of
turbulent eddies in the background turbulent convection is larger
(smaller) than the horizontal size.

(e). In order to determine values $ f_{\ast} ,$  $ \, {\bf
\Phi}^{\ast} $ and $ \, H_{\ast} $ in the background turbulent
convection we used balance equations (\ref{C1})-(\ref{C3}) for the
second moments (see Appendix A).

\subsection{Turbulent flux of entropy}

The procedure described in this Section allows us to determine the
Reynolds stresses and turbulent flux of entropy which are given by
Eqs.~(\ref{C4}) and~(\ref{C5}) in Appendix A, where we considered
the case $ \bec{\nabla}~\cdot~\bar {\bf U} = 0 .$ In particular,
the formula for turbulent flux of entropy reads:
\begin{eqnarray}
{\bf \Phi} - {\bf \Phi}^{\ast} &=&  [ - 5 \alpha (\bec{\nabla}
\cdot \bar {\bf U}_{\perp}) {\bf \Phi}_{_{\parallel}}^{\ast} +
(\alpha + 3/2) (\bec{\bar \omega} {\bf \times} {\bf
\Phi}_{_{\parallel}}^{\ast})
\nonumber \\
& & + 3 (\bec{\bar \omega}_{_{\parallel}} {\bf \times} {\bf
\Phi}^{\ast})] {\tau_{0} (q+1)\over 15} + (\bec{\nabla} {\bf
\times} {\bf T})
\nonumber \\
& & + ({\bf E} \cdot \bec{\nabla}) \bar{\bf U} \;, \label{C96}
\end{eqnarray}
where $ \, \bar \omega = (\bec{\nabla} {\bf \times} \bar{\bf U})_z
,$ $ \, {\bf \Phi}_{_{\parallel}}^{\ast} = \Phi_{z}^{\ast} {\bf e}
,$ $ \, \bec{\bar \omega}_{_{\parallel}} = \bar \omega_{z} {\bf e}
,$ $ \, {\bf T} = (2/5) \tau_{0} (q-2) ({\bf \Phi}^{\ast} \cdot
({\bf e} {\bf \times} \bar{\bf U})) $ and $ {\bf E} = (1/5)
\tau_{0} \{ [2 - q - 2 \alpha (q+1)/3 ] {\bf
\Phi}_{_{\parallel}}^{\ast} - 3 {\bf \Phi}^{\ast} \} .$ It is
shown below that the first and the second terms in Eq.~(\ref{C96})
are responsible for the large-scale convective wind instability in
a shear-free turbulent convection (see Section IV), while the
third term in the turbulent flux of entropy (\ref{C96}) causes the
convective shear instability (see Section V).

The turbulent flux of entropy can be obtained even from simple
symmetry reasoning. Indeed, this flux can be presented as a sum of
two terms: $ \langle s {\bf u} \rangle = \Phi^\ast_i + \beta_{ijk}
\nabla_j \bar U_{k} ,$ where ${\bf \Phi}^{\ast}$ determines the
contribution of the Kolmogorov turbulence and it is independent of
$ \nabla_i \bar U_{j} ,$ whereas the second term is proportional
to $ \nabla_i \bar U_{j} $ and describes the contribution of the
tangling turbulence. Here $ \beta_{ijk} $  is an arbitrary true
tensor and $ \bar {\bf U} $ is the mean velocity. Using the
identity $ \nabla_j \bar U_{i} = (\delta \bar {\bf U})_{ij} -
(1/2) \varepsilon_{ijk} \, \bar \omega_{k} ,$ the turbulent flux
of entropy becomes
\begin{eqnarray}
\langle s u_i \rangle = \Phi^\ast_i + \eta_{ij} \bar \omega_{j} +
(\bec{\bar \omega} {\bf \times} \bec{\delta})_i + \mu_{ijk}
(\delta \bar {\bf U})_{jk} \;, \label{AA3}
\end{eqnarray}
where $ (\delta \bar {\bf U})_{ij} = (\nabla_i \bar U_{j} +
\nabla_j \bar U_{i}) / 2 ,$ $\, \bec{\bar \omega} = \bec{\nabla}
{\bf \times} \bar {\bf U}$ is the mean vorticity, and $
\varepsilon_{ijk} $ is the fully antisymmetric Levi-Civita tensor.
In Eq.~(\ref{AA3}), $\eta_{ij}$ is a symmetric pseudo-tensor,
$\bec{\delta}$ is a true vector, $\mu_{ijk}$  is a true tensor
symmetric in the last two indexes, ${\bf \Phi} \equiv \langle s
{\bf u} \rangle $ and  ${\bf \Phi}^{\ast}$ are true vectors. The
tensors $\eta_{ij},$ $\, \mu_{ijk}$ and the vector $\bec{\delta}$
can be constructed using two vectors: ${\bf \Phi}^{\ast}$ and the
vertical unit vector  ${\bf e} .$ For example, $ \eta_{ij}=0 ,$
$\bec{\delta} = A_1 {\bf \Phi}^{\ast} + A_2 \Phi_{z}^{\ast} {\bf
e} ,$  and $ \mu_{ijk} = A_3 \Phi_{z}^{\ast} e_{ijk} + A_4
\Phi_{i}^{\ast} e_{jk} ,$ where $ A_k $ are the unknown
coefficients and $ e_{ijk} = e_{i} e_{j} e_{k} .$ This yields the
following expression for the turbulent flux of entropy in a
divergence-free mean velocity field:
\begin{eqnarray}
{\bf \Phi} &=& {\bf \Phi}^{\ast} - (A_3 + A_4) (\bec{\nabla} \cdot
\bar {\bf U}_{\perp}) {\bf \Phi}_{_{\parallel}}^{\ast}
\nonumber \\
& & + (A_1 + A_2) (\bec{\bar \omega} {\bf \times} {\bf
\Phi}_{_{\parallel}}^{\ast}) + A_1 (\bec{\bar
\omega}_{_{\parallel}} {\bf \times} {\bf \Phi}^{\ast})
\nonumber \\
& & - A_4 (\bec{\nabla} \cdot \bar {\bf U}_{\perp}) {\bf
\Phi}_{\perp}^{\ast}] \;, \label{AA4}
\end{eqnarray}
where $ \bar {\bf U} = \bar {\bf U}_{\perp} + \bar U_{z} {\bf e}
,$ $ \, {\bf \Phi}_{_{\parallel}}^{\ast} = \Phi_{z}^{\ast} {\bf e}
$ and $ \bec{\bar \omega}_{_{\parallel}} = \bar \omega_{z}  {\bf
e} .$ Equations ~(\ref{C96}) and~(\ref{AA4}) coincide if one sets
$A_1 = \tau_{0}(q + 1) / 5 ,$ $ \, A_2 = \tau_{0} (q + 1) (\alpha
- 3/2) / 15 ,$ $ \, A_3 = \tau_{0} \alpha (q + 1) / 3 ,$ and $ A_4
= 0 .$ Note that $ \, {\bf \Phi}^{\ast} = - \kappa_T \bec{\nabla}
\bar S - \tau_0 \Phi_{z}^{\ast} (d \bar {\bf U}^{(0)}(z) /dz) ,$ $
\, \bar {\bf U}^{(0)}(z) $ is the imposed horizontal large-scale
flow velocity (e.g., a wind velocity).

\section{Convective wind instability in a shear-free
turbulent convection}

In this section we studied the mean-field dynamics for a
shear-free turbulent convection. We showed that under certain
conditions a large-scale instability is excited, which causes
formation of large-scale semi-organized structures in a turbulent
convection.

The mean-field dynamics is determined by Eqs.~(\ref{B3}) and
(\ref{B4}). To study the linear stage of an instability we derived
linearized equations for the small perturbations from the
equilibrium, $ \bar U_{z}^{(1)} = \bar U_{z} - \bar U_{z}^{\rm
(eq)} ,$ $ \, \bar \omega^{(1)} = \bar \omega - \bar \omega^{\rm
(eq)} $ and $ \bar S^{(1)} = \bar S - \bar S^{\rm (eq)} :$
\begin{eqnarray}
\Delta {\partial \bar U_{z}^{(1)} \over \partial t} &=&  {\partial
\over \partial z} ({\nabla}_{i} {\nabla}_{j} f_{ij}^{(1)}) -
\Delta (e_{i} {\nabla}_{j} f_{ij}^{(1)})
\nonumber \\
& & + g \Delta_{\perp} \bar S^{(1)} \;,
\label{PB3} \\
{\partial \bar \omega^{(1)} \over \partial t} &=& - ({\bf e} {\bf
\times} \bec{\nabla})_i \left[{\nabla}_{j} f_{ij}^{(1)} +
{\partial \bar U_i^{\rm (eq)} \over \partial z} \bar U_{z}^{(1)}
\right] \;,
\label{PB4} \\
{\partial \bar S^{(1)} \over \partial t} &=& - (\bec{\nabla} \cdot
{\bf \Phi}^{(1)}) - \biggl(N_{b} + {\partial \bar S^{\rm (eq)}
\over  \partial z} \biggr) \bar U_{z}^{(1)} \;,
\label{PB5}
\end{eqnarray}
where $ \, f_{ij}^{(1)} = f_{ij}  - f_{ij}^{(0)} $ and the
Reynolds stresses $ f_{ij} $ are given by Eqs.~(\ref{C4}) in
Appendix A, $ \, \Delta_{\perp} = \Delta - \partial^2 / \partial
z^2 $ $ \, {\bf N}_{b} = N_{b} {\bf e} $ and
\begin{eqnarray}
\bec{\nabla} \cdot {\bf \Phi}^{(1)} &=& - (\tau_{0} / 30) (q + 1)
\{ ({\bf \Phi}^{\ast} \cdot {\bf e}) [10 \alpha \Delta_{\perp}
\nonumber \\
& &  - (8 \alpha - 3) \Delta] \bar U_{z}^{(1)} + 6 (({\bf
\Phi}^{\ast} {\bf \times} {\bf e}) \cdot \bec{\nabla}) \bar
\omega^{(1)} \}
\nonumber \\
& &  - \kappa_{ij} {\nabla}_{i} {\nabla}_{j} \bar S^{(1)} \;,
\label{C91} \\
\kappa_{ij} &=& \tau_0 \delta_{\ast}[f_{ij}^{(0)} + (1/2) g \tau_0
\delta_{\ast} (4 - \gamma) (e_i \Phi^{\ast}_{j}
\nonumber \\
& &  + e_j \Phi^{\ast}_{i})] \; .
\label{C91C}
\end{eqnarray}
Equation~(\ref{C91C}) follows from Eqs.~(\ref{C2}) and
~(\ref{C3}).

\subsection{The growth rate of convective wind instability}

Let us consider a shear-free turbulent convection $ (\nabla_{i}
\bar U_{j}^{(0)} = 0) $ with a given vertical flux of entropy $
\Phi^{\rm (eq)}_{z} \, {\bf e} .$ We also consider an isentropic
basic reference state, i.e., we neglect terms which are
proportional to $ (N_{b} + \partial \bar S^{\rm (eq)} /
\partial z) \bar U_{z}^{(1)} $  in Eq.~(\ref{PB5}). We seek for a
solution of Eqs. (\ref{PB3})-(\ref{PB5}) in the form $ \propto
\exp(\gamma_{\rm inst} t + i {\bf K} \cdot {\bf R}) ,$ where $
{\bf K} $ is the wave vector of small perturbations and $
\gamma_{\rm inst} $ is the growth rate of the instability. Thus,
the growth rate of the instability is given by
\begin{eqnarray}
\gamma_{\rm inst} = \nu_T K^{2} A [\sqrt{1 + 4 B / A^{2}} - 1] / 2  \;,
\label{C9}
\end{eqnarray}
where
\begin{eqnarray}
A = B_{1} + B_{2} \;, \quad B = \beta X (c_{7} - c_{8} X) - B_{1}
B_{2} \;, \label{C90}
\end{eqnarray}
$ \, B_{1} = c_{1} + c_{6} X - c_{3} X^{2} ,$ $ \, B_{2} = c_{4} -
c_{5} X ,$ $ \, c_{1} = (q + 3) / 5 ,$ $ \, c_{3} = \varepsilon (q
+ 3) / 4 ,$ $ \, c_{4} = \delta_{\ast} (2 + 3 \sigma) ,$ $ \,
c_{5} = 3 \delta_{\ast} (\sigma - \varepsilon / 2) ,$ $ \, c_{6} =
\varepsilon (q + 5) / 4 ,$ $ \, c_{7} = \mu (8 \alpha - 3) / 10 ,$
$ \, c_{8} = \mu \alpha ,$ with $ \sigma = a_{\ast} (4 - \gamma)
(1 + \varepsilon / 2) ,$ $ \, \mu = 6 a_{\ast} (q + 1) (1 +
\varepsilon / 2) / \delta_{\ast} ,$ $ \, \beta = (l_{0} K)^{-2} ,$
$ \, X = \sin^{2} \theta ,$ $ \, a_{\ast} = 2 \delta_{\ast}
\Phi^{\rm (eq)}_{z} \, g \tau_0 / f_{\ast} ,$ and $ \theta $ is
the angle between $ {\bf e} $ and the wave vector $ {\bf K} $ of
small perturbations. Here we used that in equilibrium $ \Phi^{\rm
(eq)}_{z} = \Phi^{\ast}_{z} .$ When $ \beta \gg 1 $ the growth
rate of the instability is given by
\begin{eqnarray}
\gamma_{\rm inst} &\propto&  \nu_T K^{2} \, \sqrt{\mu \beta} \, |
\sin \theta | \, \biggl[\alpha - {3 \over 8} - {5 \alpha \over 4}
\sin^{2} \theta \biggr]^{1/2}
\nonumber \\
&\propto& K u_{0} \; . \label{C10}
\end{eqnarray}
Thus for large $ \beta $ the growth rate of the instability is
proportional to the wave number $ K $ and the instability occurs when
$ \alpha (5 \cos^{2} \theta - 1) > 3/2 .$ This yields two ranges for
the instability:
\begin{eqnarray}
{3 \over 2 (5 \cos^{2} \theta - 1)} < & \alpha & < 3 \;,
\label{C93}\\
- {3 \over q - 1} < & \alpha & < - {3 \over 2 (1 - 5
\cos^{2} \theta)} \;,
\label{C92}
\end{eqnarray}
where we took into account that the parameter $ \alpha $ varies in
the interval $ - 3 / (q - 1) < \alpha < 3 $ (see Appendix B). The
first range for the instability in Eq.~(\ref{C93}) is for the
angles $ 3/10 \leq \cos^{2} \theta \leq 1 $ (for $ q = 5/3 ,$ the
aspect ratio $ 0 < L_{z} / L_{\perp} < 1.53) ,$ and the second
range (\ref{C92}) for the instability corresponds to the angles $
0 \leq \cos^{2} \theta < (3 - q) / 10 $ (the aspect ratio $ 2.55 <
L_{z} / L_{\perp} < \infty ) ,$ where $ L_{z} / L_{\perp} \equiv
K_{\perp} / K_{z} = \tan \theta .$ The conditions (\ref{C93}) and
(\ref{C92}) correspond to $ \bec{\nabla} \cdot {\bf \Phi}^{(1)} <
0 .$

\begin{figure}
\centering
\includegraphics[width=8cm]{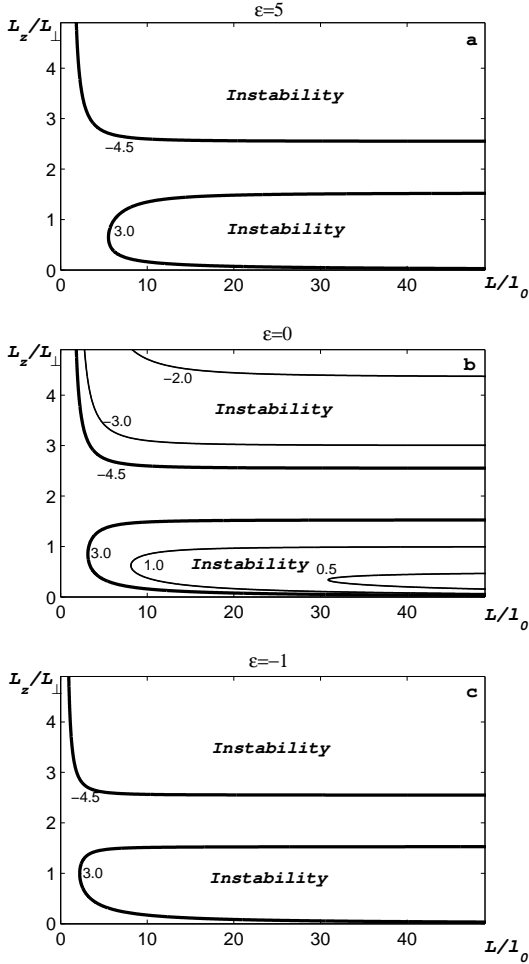}
\caption{\label{Fig1} The range of parameters $ (L_{z} /
L_{\perp}; L / l_{0}) $ for which the convective wind instability
occurs, for different values of the parameter $\alpha$: (from $ -
4.5 $ to $ 3) $ and for different values of the parameter
$\varepsilon $: a). $ \varepsilon = 5 $ ; b). $ \varepsilon = 0 $
; c). $ \varepsilon = -1 .$}
\end{figure}

Figure~\ref{Fig1} demonstrates the range of parameters $ L_{z} /
L_{\perp} $ and $ L / l_{0} $ where the instability  is excited,
for different values of the parameter $\alpha$ (from $ -4.5 $ to $
3) $ and different values of the parameter $\varepsilon = -1; \,
0; \, 5 $. Here $ L \equiv 1 / \sqrt{L_{z}^{-2} + L_{\perp}^{-2}}
$ and we assumed that $ a_\ast = 1 .$ The threshold of the
instability $ L_{\rm cr} $ depends on the parameter $\varepsilon
.$ For example, for $ \alpha = 3 $ the threshold of the
instability $ L_{\rm cr} $ varies from $3 l_{0}$ to $7 l_{0}$
(when $ \varepsilon $ changes from $-1$ to $ 5). $ The negative
(positive) degree of anisotropy $ \varepsilon $ of turbulent
velocity field corresponds to that the vertical size of turbulent
eddies in the background turbulent convection is larger (smaller)
than the horizontal size. The reason for the increase of the range
of instability with the decrease of the degree of anisotropy $
\varepsilon $ is that the rate of dissipation of the kinetic
energy of the mean velocity field decreases with decrease of $
\varepsilon $ and it causes decrease of the threshold of the
instability. The instability does not occur when $ 1.53 < L_{z} /
L_{\perp} < 2.55 $ for all $ \varepsilon .$

\begin{figure}
\centering
\includegraphics[width=8cm]{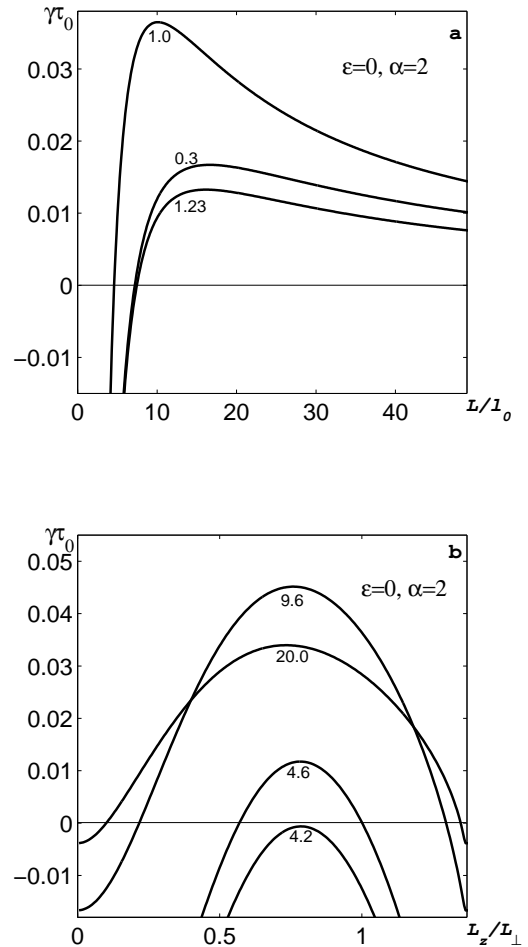}
\caption{\label{Fig2} The growth rate of the convective wind
instability as functions of: a). $ L / l_{0} $ (for different
values of parameter $ L_{z} / L_{\perp} = 0.3; \, 1; \, 1.23); $
and b). $ L_{z} / L_{\perp} $ (for different values of parameter $
L / l_{0} = 4.2; \, 4.6; \, 9.6; \, 20) $ for $ \varepsilon = 0 $
and $ \alpha = 2$.}
\end{figure}

Figure~\ref{Fig2} shows the growth rate of the instability as
function of the parameter $ L / l_{0} $ (FIG. 2a) and of the
parameter $ L_{z} / L_{\perp} $ (FIG. 2b) for $ \varepsilon = 0 $
and $ \alpha = 2$ (the first range of the instability). This range
of the instability corresponds to the ''pancake" thermal
structures of the background turbulent convection $ (l_{z} /
l_{\perp} \approx 2/3 $ for $ \alpha = 2). $ The maximum of the
growth rate of the instability $ (\gamma_{\rm max} \approx 0.045
\, \tau_{0}^{-1}) $ reaches at the scale of perturbations $ L_{m}
\approx 9.4 \, l_{0}$ (for $ L_{z} / L_{\perp} \approx 0.76). $ In
this case the threshold of the instability $ L_{\rm cr} \approx
4.2 \, l_{0} .$

\begin{figure}
\centering
\includegraphics[width=8cm]{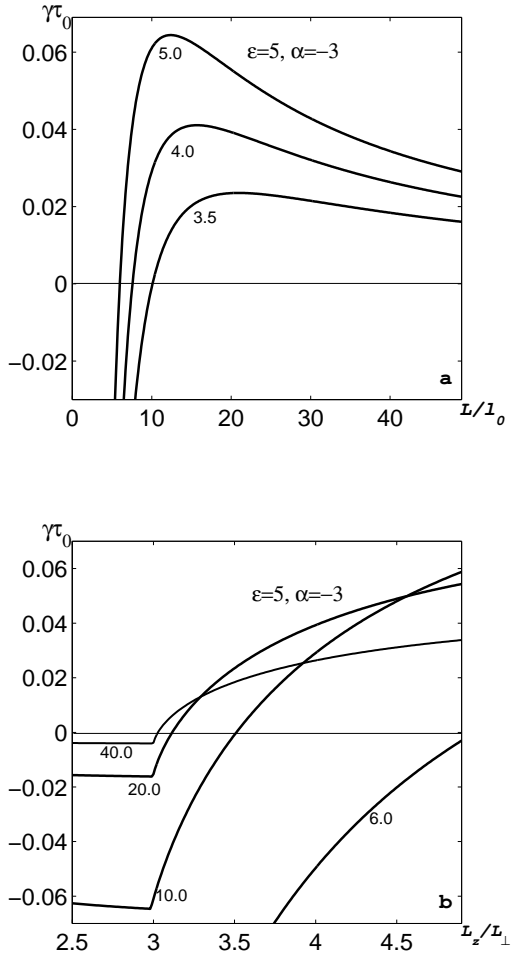}
\caption{\label{Fig3} The growth rate of the convective wind
instability as functions of: a). $ L / l_{0} $ (for different
values of parameter $ L_{z} / L_{\perp} = 3.5; \, 4; \, 5); $ and
b). $ L_{z} / L_{\perp} $ (for different values of parameter $ L /
l_{0} = 6; \, 10; \, 20; \, 40); $ for $ \varepsilon = 5 $ and $
\alpha = -3 .$}
\end{figure}

Figure~\ref{Fig3} demonstrates the growth rate for the second
range of the instability $ (\alpha = - 3). $ Note that this range
of the instability corresponds to the thermal structures of the
background turbulent convection in the form of columns $ (l_{z} /
l_{\perp} \approx 2 $ for $ \alpha = - 3). $ In contrast to the
first range of the instability, the growth rate increases with $
L_{z} / L_{\perp} $ in the whole second range of the instability
(see FIG. 3b).

\subsection{Mechanism of the convective wind instability}

The convective wind instability results in formation of
large-scale semi-organized structures in the form of cells
(convective wind) in turbulent convection. The mechanism of the
convective wind instability, associated with the first term $ {\bf
\Phi} \propto - \tau_0 \alpha (\bec{\nabla} \cdot \bar {\bf
U}_{\perp}) {\bf \Phi}_{_{\parallel}}^{\ast} $ in the expression
for the turbulent flux of entropy [see Eq.~(\ref{C96})], in the
shear-free turbulent convection at $ \alpha > 0 $ is as follows.
Perturbations of the vertical velocity $ \bar U_{z} $ with $
\partial \bar U_{z} / \partial z > 0 $ have negative divergence
of the horizontal velocity , i.e., $ {\rm div} \, \bar {\bf
U}_{\perp} < 0 $ (provided that $ {\rm div} \, \bar {\bf U}
\approx 0). $ This results in  the vertical turbulent flux of
entropy $ \tilde \Phi_{z} \propto - {\rm div} \, \bar {\bf
U}_{\perp} $, and it causes an increase of the mean entropy $
(\bar S \propto \beta^{-1/2} \, \bar U_{z} \Phi^{\ast} / u_{0}^2)
$ [see Eqs.~(\ref{PB5})-(\ref{C91}) and~(\ref{C10})].

\begin{figure}
\centering
\includegraphics[width=8cm]{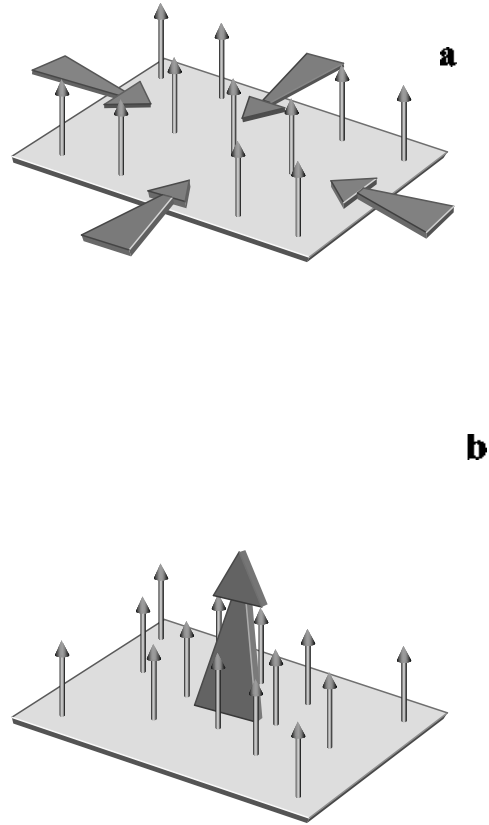}
\caption{\label{Fig4} The effect of a nonzero $ {\rm div} \, \bar
{\bf U}_{\perp} $ which causes a redistribution of the vertical
turbulent flux of the entropy and results in a formation of a
large-scale circulation of the velocity field. Fluid flow with $
{\rm div} \, \bar {\bf U}_{\perp} < 0 $ (a) produces regions with
vertical fluxes of entropy and vertical fluid flow (b) in these
regions.}
\end{figure}

On the other hand, the increase of the the mean entropy increases
the buoyancy force $ \propto g \bar S $ and results in the
increase of the vertical velocity $ \bar U_{z} \propto \tau_0
\beta^{1/2} \, g \bar S $ and excitation of the large-scale
instability [see Eqs.~(\ref{PB3}) and~(\ref{C10})]. Similar
phenomenon occurs in the regions with $ \partial \bar U_{z} /
\partial z < 0 $ whereby $ {\rm div} \, \bar {\bf U}_{\perp} > 0 .$
This causes a downward flux of the entropy and the decrease of the
mean entropy. The latter enhances the downward flow and results in
the instability which causes formation of a large-scale
semi-organized convective wind structure. Thus, nonzero $ {\rm
div} \, \bar {\bf U}_{\perp} $ causes redistribution of the
vertical turbulent flux of entropy and formation of regions with
large vertical fluxes of entropy (see FIG.~\ref{Fig4}). This
results in a formation of a large-scale circulation of the
velocity field. This mechanism determines the first range for the
instability.

The large-scale circulation of the velocity field causes a nonzero
mean vorticity $ \bar \omega ,$ and the second term [proportional
to $ (\alpha + 3/2) (\bec{\bar \omega} {\bf \times} {\bf
\Phi}_{_{\parallel}}^{\ast})] $ in the turbulent flux of the
entropy (\ref{C96}) is responsible for a formation of a horizontal
turbulent flux of the entropy. This causes a decrease of the
growth rate  of the convective wind instability (for $ \alpha >
0), $ because it decreases the mean entropy $ \bar S $ in the
regions with $ \partial \bar U_{z} / \partial z > 0 .$ The net
effect is determined by a competition between these effects which
are described by the first and the second terms in the turbulent
flux of the entropy (\ref{C96}). The latter determines a lower
positive limit $ \alpha_{\rm min} = 3/8 $ of the parameter $
\alpha .$

\begin{figure}
\centering
\includegraphics[width=8cm]{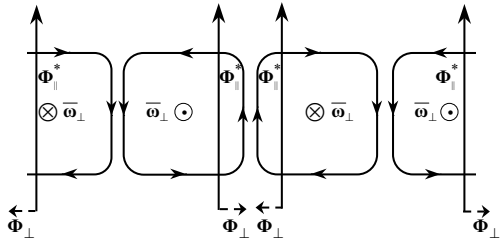}
\caption{\label{Fig5} The effect of a nonzero $ \bec{\bar \omega}
{\bf \times} {\bf \Phi}_{_{\parallel}}^{\ast} $ which induces the
horizontal flux of the mean entropy $ {\bf \Phi}_{y} $ and causes
increase (decrease) the mean entropy in the regions with upward
(downward) fluid flow when $ \alpha < - 3/2 .$}
\end{figure}

When $ \alpha < - 3/2 $ the signs of the first and second terms in
the expression~(\ref{C96}) for the turbulent flux of entropy
change. Thus, another mechanism of the convective wind instability
is associated with the second term in the expression~(\ref{C96})
for the turbulent flux of entropy when $ \alpha < - 3/2 .$ This
term describes the horizontal flux of the mean entropy $ {\bf
\Phi}_{y} \propto \tau_0 (\alpha + 3/2) (\bec{\bar \omega} {\bf
\times} {\bf \Phi}_{_{\parallel}}^{\ast}) .$ The latter results in
the increase (decrease) of the mean entropy in the regions with
upward (downward) fluid flows (see FIG.~\ref{Fig5}). On the other
hand, the increase of the mean entropy results in the increase of
the buoyancy force, the mean vertical velocity $ \bar U_{z} $  and
the mean vorticity $ \bec{\bar \omega} .$ The latter amplifies the
horizontal turbulent flux of entropy $ {\bf \Phi}_{y} $ and causes
the large-scale convective wind instability. This mechanism
determines the second range for the convective wind instability.
The first term in the turbulent flux of entropy at $ \alpha < 0 $
causes  a decrease of the growth rate of the instability because,
when $ \alpha < 0 ,$ it implies a downward turbulent flux of
entropy in the upward flow. This decreases both, the mean entropy
and the buoyancy force. Note that, when $ \alpha < - 3/2 ,$ the
thermal structure of the background turbulence has the form of a
thermal column or jets: $ l_{z} / l_{\perp} > 3.34 .$ Even for $
\alpha < 0 ,$ the ratio $ l_{z} / l_{\perp} > 1.54 .$

\section{Convective shear instability}

Let us consider turbulent convection with a linear shear $ \bar
{\bf U}^{(0)} = (\lambda / \tau_{0}) \, z \, {\bf e}_{y} $ and a
nonzero vertical flux of entropy $ {\bf \Phi} = \Phi^{\rm
(eq)}_{z} \, {\bf e} ,$ where $ \lambda $ is dimensionless
parameter which characterizes the shear. We also consider an
isentropic basic reference state, i.e., we neglected a term which
is proportional to $ (N_{b} + \partial \bar S^{\rm (eq)} /
\partial z) \bar U_{z}^{(1)} $ in Eq.~(\ref{PB5}). We seek for
a solution of Eqs. (\ref{PB3})-(\ref{PB5}) in the form $ \bar {\bf
U}^{(1)} = \bar {\bf V} \exp(\gamma_{\rm inst} t) \cos(\Omega t +
{\bf K} \cdot {\bf R}) .$ Here, for simplicity, we study the case
$ {\bf K} \cdot \bar{\bf U}^{\rm (eq)} = 0 .$

\subsection{The growth rate of convective shear instability}

Using a procedure similar to that employed for the analysis of the
convective wind instability we found that the growth rate of the
convective shear instability is determined by a cubic equation
\begin{eqnarray}
(\tilde \gamma + B_{3}) (\tilde \gamma^{2} + A \tilde \gamma - B)
+ 8 \beta^{2} \gamma_{0}^{3} = 0 \;, \label{C32}
\end{eqnarray}
where $ \tilde \gamma = (\gamma_{\rm inst} + i \Omega) / \nu_T
K^{2} ,$ $ \gamma_{0} = (1/2) c_{9}^{1/3} (\lambda X)^{2/3} ,$ $
\, c_{9} = 18 \mu b_{\ast} / 5 ,$ $ \, b_{\ast} = - \Phi^{\rm
(eq)}_{y} (1 + \varepsilon / 2) / ( \lambda \, \Phi^{\rm
(eq)}_{z}) $ and $ B_{3} = c_{1} + c_{2} X ,$ $ \, c_{2} =
\varepsilon (q + 1) / 4 .$ The growth rate of the instability for
$ \beta \gg 1 $ is given by
\begin{eqnarray}
\gamma_{\rm inst} \simeq \nu_T K^{2} \biggl(\beta^{2/3} \,
\gamma_{0} + \beta^{1/3} \, {C \over 12 \gamma_{0}} \biggr) \;,
\label{C18}
\end{eqnarray}
where $ C = X (c_{7} - c_{8} X) .$ The instability results in
generation of  the convective shear waves with the frequency
\begin{eqnarray}
\Omega \simeq \sqrt{3} \nu_T K^{2} \biggl(\beta^{2/3} \,
\gamma_{0} - \beta^{1/3} \, {C \over 12 \gamma_{0}} \biggr) \;,
\label{C19}
\end{eqnarray}
The flow in the convective shear wave has a nonzero hydrodynamic
helicity
\begin{eqnarray}
\chi \equiv \bar {\bf V} \cdot (\bec{\nabla} {\bf \times} \bar
{\bf V}) = - {2 \lambda \Omega K_{x} \bar {\bf V}^{2} \over
\tau_{0} (\Omega^{2} + \gamma_{\rm inst}^{2})} \; .
\label{C20}
\end{eqnarray}
Therefore, for $ \lambda > 0 $ the mode with $  K_{x} > 0 $ has a
negative helicity and the mode with $  K_{x} < 0 $ has a positive
helicity.

\begin{figure}
\centering
\includegraphics[width=8cm]{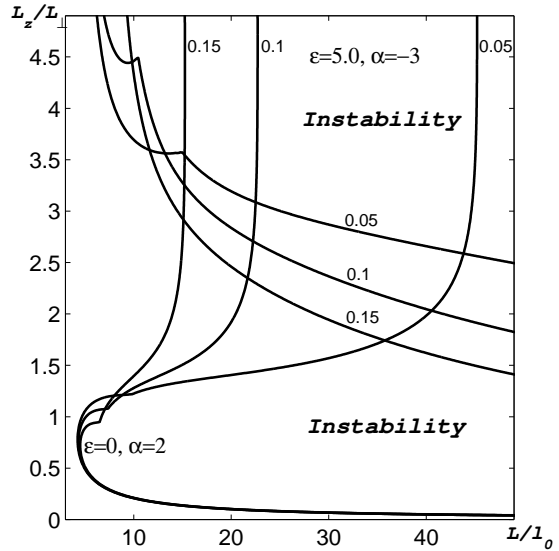}
\caption{\label{Fig6} The range of parameters $ (L_{z} /
L_{\perp}; L / l_{0}) $ for which the convective shear instability
occurs, for $\alpha = 2 ,$ $ \, \varepsilon = 0 $ and $\alpha = -3
,$ $ \, \varepsilon = 5 $ for different values of the shear $
\lambda = 0.05; \, 0.1; \, 0.15 .$}
\end{figure}

Figure~\ref{Fig6} shows the range of parameters $ L_{z} /
L_{\perp} $ and $ L / l_{0} $ where the convective shear
instability occurs, for $\alpha = 2 ,$ $ \, \varepsilon = 0 $ and
for different values of the shear $ \lambda = 0.05; \, 0.1; \,
0.15 .$ There are two ranges for the instability. However, even a
small shear causes an overlapping of the two ranges for the
instability and the increase of shear $ (\lambda) $ promotes the
convective shear instability.

\begin{figure}
\centering
\includegraphics[width=8cm]{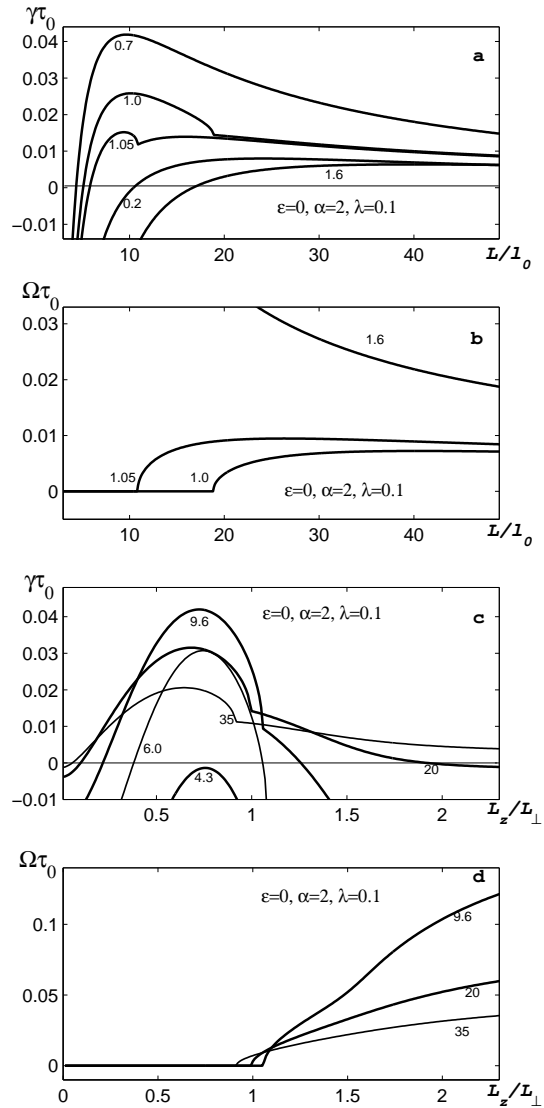}
\caption{\label{Fig7} The growth rates of the convective shear
instability [a) and c)] and the frequencies of the generated
convective shear waves [b) and d)] for the first $ (\alpha = 2) $
range of the instability and for $ \varepsilon = 0 .$
Corresponding dependencies on the parameter $ L/ l_{0} $ are given
for different $ L_{z} / L_{\perp} $ and visa versa.}
\end{figure}

\begin{figure}
\centering
\includegraphics[width=8cm]{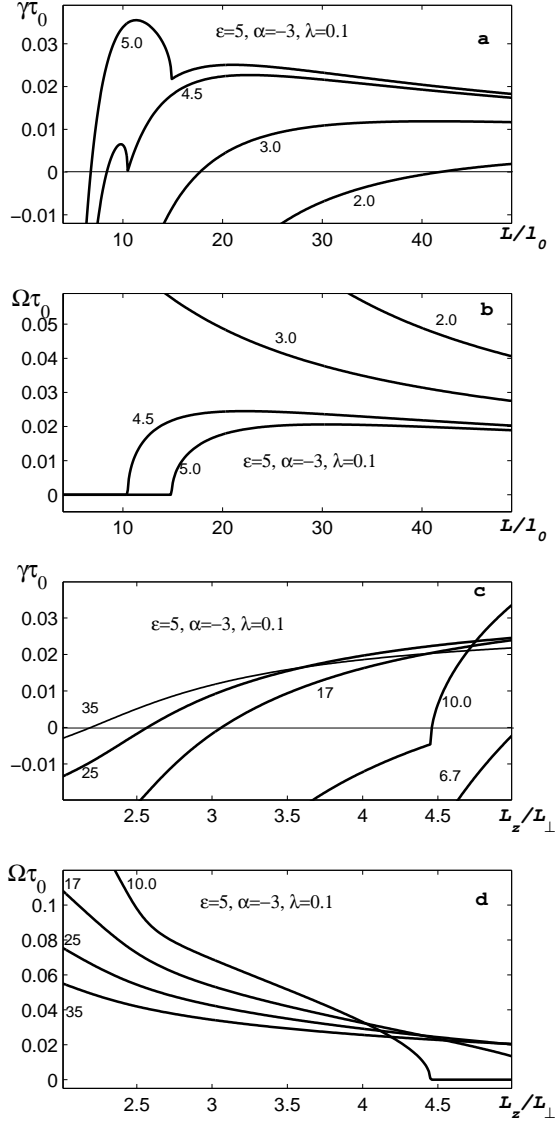}
\caption{\label{Fig8} The growth rates of the convective shear
instability [a) and c)] and the frequencies of the generated
convective shear waves [b) and d)] for the second $ (\alpha = -3)
$ range of the instability  and for $ \varepsilon = 0 .$
Corresponding dependencies on the parameter $ L/ l_{0} $ are given
for different $ L_{z} / L_{\perp} $ and visa versa.}
\end{figure}

Figures~\ref{Fig7} and~\ref{Fig8} demonstrate the growth rates of
the convective shear instability and the frequencies of the
generated convective shear waves for the first $ (\alpha = 2) $
and the second $ (\alpha = -3) $ ranges of the instability. The
curves in FIGS. \ref{Fig6}-\ref{Fig8} have a point $ L_{\ast} $
whereby the first derivative $ d \gamma_{\rm inst} / d K $ has a
singularity. At this point there is a bifurcation which is
illustrated in FIGS.~\ref{Fig7} and~\ref{Fig8}. The growth rate of
the convective shear instability is determined by the cubic
algebraic equation (\ref{C32}). Before the bifurcation point $ (L
< L_{\ast}) $ the cubic equation has three real roots (which
corresponds to aperiodic instability). After the bifurcation point
$ (L > L_{\ast}) $ the cubic equation has one real and two complex
conjugate roots. When $ L > L_{\ast} $ the convective shear waves
are generated. When the parameter $ L_{z} / L_{\perp} $ increases,
the value $ L_{\ast} $ decreases. When $ L_{z} > L_{\perp} ,$ the
bifurcation point $ L_{\ast} < L_{\rm cr} .$ For a given parameter
$ L/ l_{0} $ there are the lower and the upper bounds for  the
parameter $ L_{z} / L_{\perp} $ when the convective shear
instability occurs. For large enough parameter $ L = L_{u} $ the
upper limit of the range of the instability does not exist, e.g.,
for $ \lambda = 0.05 $ the parameter  $ L_{u} = 47 l_{0} $ and for
$ \lambda = 0.15 $ the parameter  $ L_{u} = 13 l_{0} .$

Note that when $ L < L_{\ast} $ the convective shear waves are not
generated and the properties of the convective shear instability
are similar to that  of the convective wind instability (compare
FIG. 2b and the curve for $ L/ l_{0} = 6 $ in FIG. 8c). However
for $ L > L_{\ast} $ these two instabilities are totally
different. The properties of the convective shear instability in
the first and in the second ranges of the instability are
different. In particular, in the second range of the convective
shear instability the growth rate monotonically increases, and the
frequency of the generated convective shear waves decreases with
the parameter $ L_{z} / L_{\perp} .$

\subsection{Mechanism of convective shear instability}

\begin{figure}
\centering
\includegraphics[width=8cm]{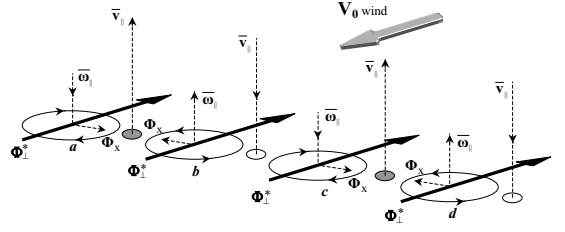}
\caption{\label{Fig9} The effect of a nonzero $ \bec{\bar
\omega}_{_{\parallel}} {\bf \times} {\bf \Phi}^{\ast} $ which
causes a redistribution of the horizontal turbulent flux of the
entropy. For two vortices ("a" and "b") with opposite directions
of the vorticity $ \bec{\bar \omega}_{_{\parallel}} ,$ the
turbulent flux of entropy is directed towards the boundary between
the vortices. The latter increases the mean entropy between the
vortices ("a" and "b"). Similarly, the mean entropy between the
vorticies "b" and "c" decreases.}
\end{figure}

The mechanism of the convective shear instability associated with
the last term in the expression~(\ref{C96}) for the turbulent flux
of entropy $ [{\bf \Phi} \propto \tau_0  (\bec{\bar
\omega}_{_{\parallel}} {\bf \times} {\bf \Phi}^{\ast})] $ is as
follows. The vorticity perturbations $ \bar \omega \equiv
(\bec{\nabla} {\bf \times} \bar {\bf U})_{z} $ generate
perturbations of entropy: $ \bar S \propto \beta^{-1/6} \,
(\tau_0/ u_0) {\Phi}^{\ast}_y \lambda^{-2/3} \exp(i \pi / 6) \bar
\omega .$ Indeed, consider two vortices (say, "a" and "b" in
FIG.~\ref{Fig9}) with the opposite directions of the vorticity $
\bec{\bar \omega}_{_{\parallel}} .$ The turbulent flux of entropy
is directed towards the boundary between the vortices. The latter
increases the mean entropy between the vortices ("a" and "b").

Similarly, the mean entropy between the vorticies "b" and "c"
decreases (see FIG.~\ref{Fig9}). Such redistribution of the mean
entropy causes increase (decrease) of the buoyancy force and
formation of upward (downward) flows between the vortices "a" and
"b" ("b" and "c"): $ \bar U_{z} \propto \beta^{1/3} \lambda^{-2/3}
g \tau_0 \, \exp(-i \pi / 3) \, \bar S .$ Finally, the vertical
flows generate vorticity $ \bar \omega \propto - \beta^{-1/6} \,
\lambda^{1/3} \exp(i \pi / 6) \bar U_{z} / l_0 ,$ etc. This
results in the excitation of the instability with the growth rate
$ \gamma_{\rm inst} \propto K^{2/3} $ and generation of the
convective shear waves with the frequency $ \Omega \propto K^{2/3}
\, .$ For perturbations with $ K_{x} = 0 $ the convective shear
instability does nor occur. However, for these perturbations with
$ K_{x} = 0 $ the convective wind instability can be excited (see
Section III), and it is not accompanied by the generation of the
convective shear waves. We considered here a linear shear for
simplicity. The equilibrium is also possible for a quadratic
shear, {\em i.e.}, when $ \bar {\bf U}^{(0)} = \tilde \lambda
z^{2} {\bf e}_{y} .$

\section{Discussion}

The "convective wind theory" of turbulent sheared convection is
proposed. The developed theory predicts the convective wind
instability in a shear-free turbulent convection. This instability
causes formation of large-scale semi-organized fluid motions
(convective wind) in the form of cells. Spatial characteristics of
these motions, such as the minimum size of the growing
perturbations and the size of perturbations with the maximum
growth rate, are determined.

This study predicts also the existence of the convective shear
instability in the sheared turbulent convection. This instability
causes formation of large-scale semi-organized fluid motions in
the form of rolls (sometimes visualized as the boundary layer
cloud streets). These motions can exist in the form of generated
convective shear waves, which have a nonzero hydrodynamic
helicity. Increase of shear promotes excitation of the convective
shear instability.

The proposed here theory of turbulent sheared convection
distinguishes between the "true turbulence", corresponding to the
small-scale part of the spectrum, and the "convective wind"
comprising of large-scale semi-organized motions caused by the
inverse energy cascade through large-scale instabilities. The true
turbulence in its turn consists of the two parts: the familiar
"Kolmogorov-cascade turbulence" and an essentially anisotropic
"tangling turbulence" caused by tangling of the mean-velocity
gradients with the Kolmogorov-type turbulence. These two types of
turbulent motions overlap in the maximum-scale part of the
spectrum. The tangling turbulence does not exhibit any direct
energy cascade.

\begin{figure}
\centering
\includegraphics[width=8cm]{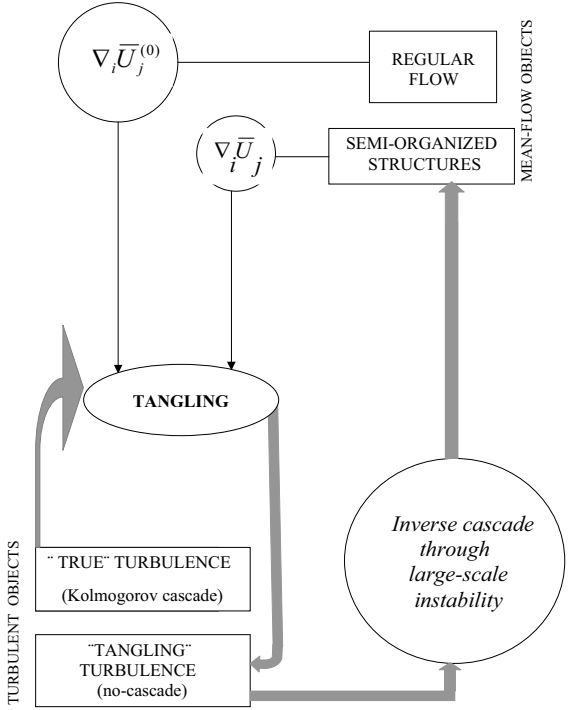}
\caption{\label{Fig10} Scheme of interactions between turbulent
and mean-flow objects which cause a large-scale instability.}
\end{figure}

It was demonstrated here that the characteristic length and time
scales of the convective wind motions are much larger than the
true-turbulence scales. This justifies separation of scales which
is required for the existence of these two types of motions. It is
proposed that the term turbulence (or true turbulence) be kept
only for the Kolmogorov and tangling turbulence part of the
spectrum. This concept implies that the convective wind (as well
as semi-organized motions in other very high Reynolds number
flows) should not be confused with the true turbulence. The
diagram of interactions between turbulent and mean flow objects
which cause the large-scale instability and formation of
semi-organized structures is shown in FIG.~\ref{Fig10}.

Now let us compare the obtained results with the properties of
semi-organized structures observed in the atmospheric convective
boundary layer. The semi-organized structures are observed in the
form of rolls (cloud streets) or three-dimensional convective
cells (cloud cells). Rolls usually align along or at angles of up
to $10^{\circ}$ with the mean horizontal wind of the convective
layer, with lengths from $20$ to $200$ km, widths from $2$ to $10$
km, and convective depths from $2$ to $3$ km \cite{AZ96}. The
typical value of the aspect ratio $ L_{z} / L_{\perp} \approx 0.14
- 1 .$ The ratio of the minimal size of structures to the maximum
scale of turbulent motions is $ L / l_{0} = 10 - 100 .$ The
characteristic life time of rolls varies from $1$ to $72$ hours
\cite{EB93}. Rolls may occur over both, water surface and land
surfaces. The suggested theory predicts the following parameters
of the convective rolls: the aspect ratio $ L_{z} / L_{\perp} $
ranges from very small to  $ 1 ,$ and $ L / l_{0} = 10 - 100 .$
The characteristic time of formation of the rolls $ \sim \tau_0 /
\gamma_{\rm inst} $ varies from $1$ to $3$ hours. The life time of
the convective rolls is determined by a nonlinear evolution of the
convective shear instability. The latter is a subject of a
separate ongoing study.

Convective cells may be divided into two types: open and closed.
Open-cell circulation has downward motion and clear sky in the
cell center, surrounded by cloud associated with upward motion.
Closed cells have the opposite circulation \cite{AZ96}. Both types
of cells have diameters ranging from $10$ to $40$ km and aspect
ratios $ L_{z} / L_{\perp} \approx  0.05 - 1 ,$ and both occur in
a convective boundary layer with a depth of about $1$ to $3$ km.
The ratio of the minimum size of structures to the maximum scale
of turbulent motions is $ L / l_{0} = 5 - 20 .$ The developed
theory predicts the following parameters of the convective cells:
the aspect ratio $ L_{z} / L_{\perp} $ ranges from very small to $
1 ,$ and $ L / l_{0} = 5 - 15 .$ The characteristic time of
formation of the convective cells $ \sim \tau_0 / \gamma_{\rm
inst} $ varies from $1$ to $3$ hours. Therefore the predictions of
the developed theory are in a good agreement with observations of
the semi-organized structures in the atmospheric convective
boundary layer. Moreover, the typical temporal and spatial scales
of structures are always much larger then the turbulence scales.
This justifies separation of scales which was assumed in the
suggested in the theory.

\begin{acknowledgments}
We have benefited from valuable suggestions made by Arkady
Tsinober. The authors acknowledge useful discussions with Erland
K\"{a}llen and Branko Grisogono at seminar at the Meteorological
Institute of Stockholm University. This work was partially
supported by The German-Israeli Project Cooperation (DIP)
administrated by the Federal Ministry of Education and Research
(BMBF), by INTAS Program Foundation (Grants No. 00-0309 and No.
99-348), by the SIDA Project SRP-2000-036, and by the Swedish
Institute Project 2570/2002 (381/N34).
\end{acknowledgments}

\appendix

\section{Derivations of expressions  for the Reynolds stresses
and turbulent flux of entropy}

Equations (\ref{B1}) and (\ref{B2}) yield the following
conservation equations for the kinetic energy $ W_{v} = \rho_0
{\bf v}^{2} / 2 ,$ for $ W_{S} = \rho_0 S^{2} / 2 $ and for $ {\bf
W}^{\Phi} = \rho_0 S {\bf v} :$
\begin{eqnarray}
{\partial  W_{v}  \over \partial t} + \bec{\nabla} \cdot {\bf
F}_{v} &=& I_{v} - D_{v} \;,
\label{B16} \\
{\partial  W_{S}  \over \partial t} + \bec{\nabla} \cdot {\bf
F}_{S} &=& I_{S} - D_{S} \;,
\label{B17} \\
{\partial  W_{i}^{\Phi}  \over \partial t} + \bec{\nabla}_{j} {\bf
F}_{ij}^{\Phi} &=& I_{i}^{\Phi} - D_{i}^{\Phi} \;, \label{B18}
\end{eqnarray}
where $ I_{v} = - \rho_0 ({\bf v} \cdot {\bf g}) S ,$ $ \, I_{S} =
- I_{v} \Omega_{b}^{2} / g^{2} ,$ and $ {\bf I}^{\Phi} = - \rho_0
[{\bf v} ({\bf v}~\cdot~{\bf N}_{b}) + S^{2} {\bf g}] + (P /
\rho_0) \bec{\nabla} (S \rho_0) $ are the source terms in these
equations, $ D_{v} = - \rho_0 ({\bf v} \cdot {\bf f}_{\nu}) ,$ $
\, D_{S} = \rho_0 S \, (\bec{\nabla} \cdot {\bf F}_{\kappa}) $ and
$ {\bf D}^{\Phi} = - \rho_0 S {\bf f}_{\nu} + (\rho_0 / T_{0})
{\bf v} (\bec{\nabla} \cdot {\bf F}_{\kappa}) $ are the
dissipative terms, $ {\bf F}_{v} = {\bf v} \, (W_{v} + P) ,$ $ \,
{\bf F}_{S} = {\bf v} W_{S} $ and $ {\bf F}_{ij}^{\Phi} = \rho_0 S
v_{i} v_{j} + S  P \delta_{ij} $ are the fluxes. Equations
(\ref{B16}) and (\ref{B17}) yield conservation equation for $
W_{E} = W_{v} \Omega_{b}^{2} / g^{2} + W_{S}$
\begin{eqnarray}
{\partial  W_{E}  \over \partial t} + \bec{\nabla} \cdot {\bf
F}_{E} &=& - D_{E} \;, \label{B19}
\end{eqnarray}
where $ D_{E} = D_{v} \Omega_{b}^{2} / g^{2} + D_{S} $ is the
dissipative term, and $ {\bf F}_{E} = {\bf F}_{v} \Omega_{b}^{2} /
g^{2} + {\bf F}_{S} $ is the flux. Equation (\ref{B19}) does not
have a source term, and this implies that without dissipation $
(D_{E} = 0) $ the value $ \int W_{E} \, dV $ is conserved, where
in the latter formula the integration is performed over the
volume. For the convection $ \Omega_{b}^{2} < 0 $ and, therefore,
$ W_{S} \approx W_{v} |\Omega_{b}^{2}| / g^{2} .$

Using Eqs. (\ref{B16})-(\ref{B18}) we derived balance equations
for the second moments. In particular, averaging Eqs.
(\ref{B16})-(\ref{B18}) over the ensemble of fluctuations and
subtracting from these equations the corresponding equations for
the mean fields: $ \rho_{0} \bar {\bf U}^{2} / 2 ,$ $ \, \rho_{0}
\bar S^{2} / 2 ,$ $ \, \rho_{0} \bar S \bar {\bf U} ,$ yields
\begin{eqnarray}
\biggl({\partial \over \partial t} &+& \bar {\bf U} \cdot
\bec{\nabla} \biggr) f_{pp} + 2 f_{ij} \nabla_{j} \bar U_{i} + 2
{\bf \Phi} {\bf \cdot} {\bf g} + {2 \over \rho_{0}} \bec{\nabla}
\cdot {\bf \Psi}_E
\nonumber \\
& & = - {f_{\ast} \over \tau_0 \delta_{\ast}} \biggl(1 +
{\varepsilon \over 2} \biggr) \;,
\label{C1} \\
\biggl({\partial \over \partial t} &+& \bar {\bf U} \cdot
\bec{\nabla}\biggr) \Phi_{i} + ({\bf \Phi} \cdot \bec{\nabla})
\bar U_{i} + f_{ij}({\bf N_{b}} + \bec{\nabla} \bar S)_{j}
\nonumber \\
& & - {1\over 2} g e_{i} (4 - \gamma) H + {1 \over \rho_{0}}
\nabla_{j} \Psi_{ij} - \biggl({\gamma P_{0} \over \rho_{0}}
\biggr) H ({\bf N_{b}})_{i}
\nonumber \\
& & = - {\Phi^{\ast}_{i} \over 2 \tau_0 \delta_{\ast}} ( 1 + {\rm
Pr}) \;,
\label{C2} \\
\biggl({\partial \over \partial t} &+& \bar {\bf U} \cdot
\bec{\nabla} \biggr) H + 2 \Phi_{j}({\bf N_{b}} + \bec{\nabla}
\bar S)_{j} = - {H_{\ast} \over \tau_0 \delta_{\ast}} \;,
\label{C3}
\end{eqnarray}
where $ {\rm Pr} = \nu / \kappa $ is the Prandtl number, $ \nu $
is the kinematic viscosity, $ {\bf \Psi}_E = \rho_{0} f_{pp} {\bf
u} / 2 + \langle p {\bf u} \rangle ,$ $ \, \Psi_{ij} = \rho_{0}
\langle s u_{i} u_{j} \rangle + \delta_{ij} (\gamma P_{0} H / 2 +
\langle s p \rangle) ,$ and we took into account that the
dissipations of energy, the flux of entropy and the second moment
of entropy $ H $ are determined by the background turbulent
convection described by Eqs. (\ref{B8})-(\ref{B12}). In derivation
of Eq.~(\ref{C2}) we used an identity $ \rho_{0} S \bec{\nabla} (P
/ \rho_{0}) \simeq \bec{\nabla}(\gamma P_{0} S^{2} ) / 2 + \gamma
P_{0} S^{2} {\bf N_{b}} + (\gamma / 2 - 1) \rho_{0} {\bf g} S^{2}
,$ and we assumed that $ S \simeq P / \gamma P_{0} ,$ {\em i.e.,}
we neglected fluctuations of density $ \rho / \rho_{0} .$
Equations (\ref{C1})-(\ref{C3}) allow us to determine $ f_{\ast}
,$ $ \, {\bf \Phi}^{\ast} $ and $ \, H_{\ast} $ in the background
turbulent convection (see below).

Using Eqs.~(\ref{B5}) and (\ref{B6}) we derived equations for the
following second moments:
\begin{eqnarray}
f_{ij}({\bf k},{\bf R}) &=& \langle u_{i}({\bf k},{\bf R})
u_{j}(-{\bf k},{\bf R}) \rangle \;, \label{A3}\\
\Phi_{i}({\bf k},{\bf R}) &=& \langle s({\bf k},{\bf R})
u_{i}(-{\bf k},{\bf R}) \rangle \;,
\label{MA3} \\
F({\bf k},{\bf R}) &=& \langle s({\bf k},{\bf R}) \omega(-{\bf
k},{\bf R}) \rangle \;,
\label{MA4}\\
G({\bf k},{\bf R}) &=& \langle \omega({\bf k},{\bf R})
\omega(-{\bf k},{\bf R}) \rangle \;,
\label{A4} \\
H({\bf k},{\bf R}) &=& \langle s({\bf k},{\bf R}) s(-{\bf k},{\bf
R}) \rangle \;, \label{A5}
\end{eqnarray}
where $ \omega \equiv (\bec{\nabla} {\bf \times} {\bf u})_z $ and
we use a two-scale approach, , {\em i.e.,} a correlation function
is written as follows
\begin{eqnarray*}
\langle u_i({\bf x}) u_j ({\bf  y}) \rangle &=& \int \langle u_i
({\bf k}_1) u_j ({\bf k}_2) \rangle  \exp[i({\bf  k}_1 {\bf \cdot}
{\bf x} \\
&& + {\bf k}_2 {\bf \cdot} {\bf y})] \,d{\bf k}_1 \, d{\bf k}_2
\\
&=& \int f_{ij}({\bf k, R}) \exp(i {\bf k} {\bf \cdot} {\bf r})
\,d {\bf k} \;,
\\
f_{ij}({\bf k, R}) &=& \int \langle u_i ({\bf k} + {\bf  K} / 2)
u_j(-{\bf k} + {\bf  K} / 2) \rangle
\\
&& \times \exp(i {\bf K} {\bf \cdot}{\bf R}) \,d {\bf K} \;
\end{eqnarray*}
(see, {\em e.g.,} \cite{RS75,KR94}), where $ {\bf R} $ and $ {\bf
K} $ correspond to the large scales, and $ {\bf r} $ and $ {\bf k}
$ to the small scales, {\em i.e.,} $ {\bf R} = ({\bf x} +  {\bf
y}) / 2  ,$ $ \quad {\bf r} = {\bf x} - {\bf y},$ $ \quad {\bf K}
= {\bf k}_1 + {\bf k}_2,$ $ \quad {\bf k} = ({\bf k}_1 - {\bf
k}_2) / 2 .$ This implies that we assumed that there exists a
separation of scales, i.e., the maximum scale of turbulent motions
$ l_0 $ is much smaller then the characteristic scale $ L $ of
inhomogeneities of the mean fields. In particular, this implies
that $ r \leq l_0 \ll R .$ Our final results showed that this
assumption is indeed valid. Now let us calculate
\begin{eqnarray}
{\partial f_{ij}({\bf k}_1,{\bf k}_2) \over \partial t} &\equiv&
\langle P_{in}({\bf k}_1) {\partial u_{n}({\bf k}_1) \over \partial t}
u_j({\bf k}_2) \rangle
\nonumber\\
&& + \langle u_i({\bf k}_1) P_{jn}({\bf k}_2) {\partial u_{n}({\bf
k}_2) \over \partial t} \rangle \;,
\label{A8} \\
{\partial \Phi_{j}({\bf k}_1,{\bf k}_2) \over \partial t} &\equiv&
\langle s({\bf k}_1) P_{jn}({\bf k}_2) {\partial u_{n}({\bf k}_2) \over
\partial t} \rangle
\nonumber\\
&& + \langle {\partial s({\bf k}_1) \over \partial t} u_j({\bf
k}_2) \rangle \;, \label{A9}
\end{eqnarray}
where we multiplied equation of motion (\ref{B5}) rewritten in
$ {\bf k} $-space by $ P_{ij}({\bf k}) = \delta_{ij} - k_{ij} $
in order to exclude the pressure term from the equation of motion.

Thus, equations for $ f_{ij}({\bf k, R}) $ and $ {\bf \Phi}({\bf
k, R}) $ read:
\begin{eqnarray}
{\partial f_{ij}({\bf k}) \over \partial t} &=& \hat I_{ijmn} f_{mn}({\bf k})
+ N_{ij}({\bf k}) \;,
\label{A6} \\
{\partial \Phi_{i}({\bf k}) \over \partial t} &=& \hat I_{ij} \Phi_{j}({\bf k})
+ M_{i}({\bf k}) \;,
\label{A7}
\end{eqnarray}
where
\begin{eqnarray}
\hat I_{ijmn} &=& 2 (k_{iq} \delta_{mp} \delta_{jn} + k_{jq}
\delta_{im} \delta_{pn}) \nabla_{p} \bar U_{q} -
\biggl(\delta_{im} \delta_{jq} \delta_{np}
\nonumber\\
&& + \delta_{iq} \delta_{jn} \delta_{mp} - \delta_{im} \delta_{jn}
k_{q} {\partial \over \partial k_{p}}\biggr) \nabla_{p} \bar U_{q}
\;,
\label{A14} \\
\hat I_{ij} &=& 2 k_{in} \nabla_{j} \bar U_{n} + \delta_{ij} k_{n}
{\partial \over \partial k_{m}} \nabla_{m} \bar U_{n} - \nabla_{j}
\bar U_{i} \;,
\label{A15} \\
N_{ij}({\bf k}) &=& g e_{m} [P_{im}({\bf k}) \Phi_{j}({\bf k})
+ P_{jm}({\bf k}) \Phi_{i}(-{\bf k})]
\nonumber\\
&& + f_{ij}^{N}({\bf k})  \;,
\label{A16} \\
M_{i}({\bf k}) &=& - f_{mi}({\bf N_{b}} + \bec{\nabla} \bar S)_{m}
+ g e_{m} P_{im}({\bf k}) H
\nonumber\\
&& + \Phi_{i}^{N}   \;, \label{A17}
\end{eqnarray}
and hereafter we consider the case with $\bec{\nabla} \cdot \bar
{\bf U}=0 $ (i.e., $\Lambda=0) .$ Here $ f_{ij}^{N} $ and $
\Phi_{i}^{N} $ are the third moments appearing due to the
nonlinear terms. Equations (\ref{A6}) and (\ref{A7}) are written
in a frame moving with a local velocity $ \bar {\bf U} $ of the
mean flow. In Eqs.~(\ref{A6})-(\ref{A17}) we neglected small terms
which are of the order of $O(\nabla^3 \bar{\bf U})$ and
$O(\nabla^2 f_{ij}; \nabla^2 \Phi_{i}) .$ Note that
Eqs.~(\ref{A6})-(\ref{A17}) do not contain the terms proportional
to $O(\nabla^2 \bar{\bf U}) .$ The first term in the RHS of
Eqs.~(\ref{A6}) and (\ref{A7}) depends on the gradients of the
mean fluid velocity $ (\nabla_{i} \bar U_{j}). $ Equations for the
second moments $ G({\bf k}) ,$ $ \, F({\bf k}) $ and $ H({\bf k})
$ read:
\begin{eqnarray}
{\partial G({\bf k}) \over \partial t} &=& ({\bf e} {\bf \times}
\tilde{\bf k}_{1})_{i} ({\bf e} {\bf \times} \tilde{\bf k}_{2})_{j}
\, {\partial f_{ij}({\bf k}) \over \partial t}  \;,
\label{A10} \\
{\partial F({\bf k}) \over \partial t} &=& - i ({\bf e} {\bf \times}
\tilde{\bf k}_{1})_{j} \, {\partial \Phi_{j}({\bf k}) \over \partial t}  \;,
\label{A11} \\
{\partial H({\bf k}) \over \partial t} &=& Q({\bf k}) \;,
\label{A12}
\end{eqnarray}
where $ Q({\bf k}) = - 2 {\bf \Phi}({\bf k}) \cdot ({\bf N}_{b} +
\bec{\nabla} \bar S) + H_{N} ,$ and $ H_{N} $ is the third moment
appearing due to  the nonlinear terms, $ \tilde{\bf k}_{1} = {\bf
k} - (i / 2) \bec{\nabla} ,$ $ \, \tilde{\bf k}_{2} = {\bf k} + (i
/ 2) \bec{\nabla} .$ The terms $ \sim {\bf \Phi} $ in the tensor $
N_{ij}({\bf k}) $ [see Eqs.~(\ref{A6}) and (\ref{A16})] can be
considered as a stirring force for the turbulent convection. On
the other hand, the terms $ \sim ({\bf N}_{b} + \bec{\nabla} \bar
S) $ in Eqs.~(\ref{A7}), (\ref{A17}) and (\ref{A12}) are the
sources of the flux of entropy $ {\bf \Phi} $ and the second
moment of entropy $ H .$ Note that a stirring force in the
Navier-Stokes turbulence is an external parameter.

Since the equations for the second moments contain the third
moments, a problem of closure for the higher moments arises. In
this study we used the $ \tau $ approximation [see Eq.~(\ref{B7})]
which allows us to express the third moments $ f^{N}_{ij} ,$ $
{\bf \Phi}^{N} $ and $ H_{N} $ in Eqs.~(\ref{A6}), (\ref{A7}) and
(\ref{A12}) in terms of the second moments. Here we define a
background turbulent convection as the turbulent convection with
zero gradients of the mean fluid velocity $ (\nabla_{i} \bar U_{j}
= 0)]. $ The background turbulent convection is determined by the
following equations:
\begin{eqnarray}
{\partial f_{ij}^{(0)}({\bf k}) \over \partial t} &=& N_{ij}^{(0)}({\bf k}) \;,
\label{A18} \\
{\partial \Phi_{i}^{(0)}({\bf k}) \over \partial t} &=& M_{i}^{(0)}({\bf k})
\;,
\label{A19} \\
{\partial H^{(0)}({\bf k}) \over \partial t} &=& Q^{(0)}({\bf k})
\; . \label{A22}
\end{eqnarray}
A nonzero gradient of the mean fluid velocity results in
deviations from the background turbulent convection. These
deviations are determined by the following equations:
\begin{eqnarray}
{\partial (f_{ij} - f_{ij}^{(0)}) \over \partial t} &=&
\hat I_{ijmn} \, f_{mn}({\bf k}) - {f_{ij} - f_{ij}^{(0)} \over \tau(k)} \;,
\label{A20} \\
{\partial (\Phi_{i} - \Phi_{i}^{(0)}) \over \partial t} &=&
\hat I_{ij} \, \Phi_{j}({\bf k}) - {\Phi_{i} - \Phi_{i}^{(0)} \over \tau(k)} \;,
\label{A21} \\
{\partial (H - H^{(0)}) \over \partial t} &=& - {H - H^{(0)} \over
\tau(k)} \;, \label{A23}
\end{eqnarray}
where the deviations (caused by a nonzero gradients of the mean
fluid velocity) of the functions  $ N_{ij}({\bf k}) -
N_{ij}^{(0)}({\bf k}) $ and $ M_{i}({\bf k}) - M_{i}^{(0)}({\bf
k}) $ from the background state are described by the relaxation
terms: $ - (f_{ij} - f_{ij}^{(0)}) / \tau(k) $ and $ - (\Phi_{i} -
\Phi_{i}^{(0)}) / \tau(k) ,$ respectively. Similarly, the
deviation $ Q({\bf k}) - Q^{(0)}({\bf k}) $ is described by the
term $ -(H - H^{(0)}) / \tau(k) .$ Here we assumed that the
correlation time $ \tau(k) $ is independent of the gradients of
the mean fluid velocity.

Now we assume that the characteristic times of variation of the
second moments $ f_{ij}({\bf k}) ,$ $ \Phi_{i}({\bf k}) $ and $
H({\bf k}) $ are substantially larger than the correlation time $
\tau(k) $ for all turbulence scales. This allows us to determine a
stationary solution for the second moments $ f_{ij}({\bf k}) ,$ $
\Phi_{i}({\bf k}) $ and $ H({\bf k}):$
\begin{eqnarray}
f_{ij}({\bf k}) &=& f_{ij}^{(0)}({\bf k}) +
\tau(k) \hat I_{ijmn} \, f_{mn}^{(0)}({\bf k}) \;,
\label{A24} \\
\Phi_{i}({\bf k}) &=& \Phi_{i}^{(0)}({\bf k}) +
\tau(k) \hat I_{ij} \, \Phi_{j}^{(0)}({\bf k}) \;,
\label{A25} \\
H({\bf k}) &=& H^{(0)}({\bf k}) \;, \label{A26}
\end{eqnarray}
where we neglected the third and higher order spatial derivatives
of the mean velocity field $ \bar {\bf U} .$

For the integration in $ {\bf k} $-space of the second moments $
f_{ij}({\bf k}) ,$ $ \Phi_{i}({\bf k}) , \ldots ,$ $ H({\bf
k},{\bf R}) $ we have to specify a model for the background
turbulent convection. We used the model of the background
turbulent convection determined by Eqs. (\ref{B8})-(\ref{B12}).
For the integration in $ {\bf k} $-space we used identities given
in Appendix C. The integration in $ {\bf k} $-space of
Eqs.~(\ref{A24}) and (\ref{A25}) yields the following equations
for the Reynolds stresses and the turbulent flux of entropy:
\begin{eqnarray}
f_{ij} &=& f_{ij}^{(0)} - \nu_T \biggl(a_{1} \, (\nabla_{i} \bar
U_{j} + \nabla_{j} \bar U_{i}) + a_{2} \, (e_{i} \nabla_{j}
\nonumber \\
& & + e_{j} \nabla_{i}) \bar U_{z} - c_{2} \, {\partial \over
\partial z} (e_{i} \bar U_{j} + e_{j} \bar U_{i})
\nonumber \\
& & + {\partial \bar U_{z} \over \partial z} (c_{3} \, e_{ij} +
a_{3} \, \delta_{ij}) \biggr) \;,
\label{C4} \\
{\bf \Phi} &=& {\bf \Phi}^{\ast} + {\tau_{0} \over 30} ({\bf
\Phi}^{\ast} \cdot {\bf e}) \biggl( {\partial \over \partial z}
(b_{1} \, \bar U_{z} {\bf e} + b_{2} \, \bar{\bf U}) - b_{3} \,
\bec{\nabla}^{\perp} \bar U_{z} \biggr)
\nonumber \\
& & - {\tau_{0} \over 5} \biggl[2 ({\bf \Phi}^{\ast} {\bf \times}
{\bf e}) \bar \omega + 5 ({\bf \Phi}^{\ast} \cdot \bec{\nabla})
\bar{\bf U} - 2 ({\bf \Phi}^{\ast} \cdot \bec{\nabla}_{\perp})
\bar{\bf U}
\nonumber \\
& & + (3-q) ({\bf e} {\bf \times} \bec{\nabla}) ({\bf \Phi}^{\ast}
\cdot ({\bf e} {\bf \times} \bar{\bf U}))
\nonumber \\
& & - (q-1) [(({\bf \Phi}^{\ast} {\bf \times} {\bf e}) \cdot
\bec{\nabla}) ({\bf e} {\bf \times} \bar{\bf U}) \biggr] \,,
\label{C5}
\end{eqnarray}
where $ \nu_T = \tau_{0} f_{\ast} / 6 ,$ $ \, \bar \omega =
(\bec{\nabla} {\bf \times} \bar{\bf U})_z ,$ $ \, a_{1} = c_{1} +
c_{2} ,$ $ \, a_{2} = - \varepsilon (q - 1) / 4 ,$ $ \, a_{3} = -
\varepsilon (5 - q) / 4 ,$ $ \, b_{1} = (8 \alpha - 3) (q + 1) ,$
$ \, b_{2} = 3 (9 - q) - 2 \alpha (q + 1) ,$ $ \, b_{3} = (2
\alpha + 3) (q + 1) ,$ $ \, c_{1} = (q + 3) / 5 ,$ $ \, c_{2} =
\varepsilon (q + 1) / 4 ,$ $ \, c_{3} = \varepsilon (q + 3) / 4 .$

Equations~(\ref{C4}) and (\ref{C5}) imply that there are two
contributions to the Reynolds stresses and turbulent flux of
entropy which correspond to two kinds of fluctuations of the
velocity field. The first contribution is due to the Kolmogorov
turbulence with the spectrum $ (\propto k^{-q}), $ and it
corresponds to the background turbulent convection. The second
kind of fluctuations depends on gradients of the mean velocity
field and is caused by a "tangling" of gradients of the mean
velocity field by turbulent motions. The spectrum of the tangling
turbulence is $ W (k) \tau(k) \propto k^{1-2q} $ [see
Eqs.~(\ref{A24}) and (\ref{A25})]. These fluctuations describe
deviations from the background turbulent convection caused by the
gradients of the mean fluid velocity field.

Now we calculate a dissipation of the kinetic energy of the mean
flow $ \bar {\bf U} :$
\begin{eqnarray}
D_{U} \equiv - (1/2) (f_{ij} - f_{ij}^{(0)}) (\nabla_{i} \bar
U_{j} + \nabla_{j} \bar U_{i}) \;, \label{C34}
\end{eqnarray}
using a general form of the velocity field $ \bar U_{i} = \bar
V_{i}(t,{\bf K}) \exp(i {\bf K} \cdot {\bf R}) ,$ where
\begin{eqnarray}
\bar V_{i}(t,{\bf K}) &=& \biggl({K \over K_{\perp}}\biggr)^{2}
[P_{ij}({\bf K}) e_{j} \bar V_{z}(t,{\bf K})
\nonumber \\
& & - i K^{-2} ({\bf e} {\bf \times} {\bf K})_{i} \, \tilde
\omega(t,{\bf K})] \; , \label{C35}
\end{eqnarray}
and $\tilde \omega = (\bec{\nabla} {\bf \times} \bar   {\bf
V})_{z} .$ The result is given by
\begin{eqnarray}
D_{U}(t,{\bf K}) &=& \nu_T \biggl\{b_{4} \, \biggl({K \over
K_{\perp}} \biggr)^{2} [K^{2} \bar V_{z}^{2}(t,{\bf K}) + \tilde
\omega^{2}(t,{\bf K})]
\nonumber \\
&& + b_{5} \, K^{2} \bar V_{z}^{2}(t,{\bf K}) \biggr\} \;,
\label{C36}
\end{eqnarray}
where $ b_{4} = c_{1} + c_{2} \sin^{2} \theta $ and $ b_{5} =
a_{2} + c_{3} \cos^{2} \theta .$ The function $ D_{U}(t,{\bf K}) $
must be positive for statistically stationary small-scale
turbulence. The latter is valid when $\varepsilon$ satisfies
condition~(\ref{C37}).

\section{The model of the background turbulent convection}

A simple approximate model for the three-dimensional isotropic
Navier-Stokes turbulence is described by a two-point correlation
function of the velocity field $ f_{ij}(t,{\bf x},{\bf y}) =
\langle u_i(t,{\bf x}) u_j(t,{\bf y}) \rangle $ with the
Kolmogorov spectrum $ W(k) \propto k^{-q} $ and $ q = 5/3 .$ The
turbulent convection is determined not only by the turbulent
velocity field $ {\bf u}(t,{\bf x}) $ but by the fluctuations of
the entropy $ s(t,{\bf x}) .$ This implies that for the
description of the turbulent convection one needs additional
correlation functions, e.g., the turbulent flux of entropy $ {\bf
\Phi}_{i}(t,{\bf x},{\bf y}) = \langle s(t,{\bf x}) u_i(t,{\bf y})
\rangle $ and the second moment of the entropy fluctuations $
H(t,{\bf x},{\bf y}) = \langle s(t,{\bf x}) s(t,{\bf y}) \rangle
.$ Note also that the turbulent convection is anisotropic.

Let us derive Eqs.~(\ref{B8}) and (\ref{B9}) for the correlation
functions $ f_{ij} $ and $ {\bf \Phi}_{i} .$ To this end, the
velocity $ {\bf u}_{\perp} $ is written as a sum of the vortical
and the potential components, {\em i.e.,} $ {\bf u}_{\perp} =
\bec{\nabla} {\bf \times} (\tilde C {\bf e}) +
\bec{\nabla}_{\perp} \tilde \phi ,$ where $ \omega \equiv
(\bec{\nabla} {\bf \times} {\bf u})_z = - \Delta_{\perp} \tilde C
,$ $ \, \Delta_{\perp} \tilde \phi = - \partial u_{z} /
\partial z ,$ $ \, \bec{\nabla}_{\perp} = \bec{\nabla} - {\bf e}
({\bf e} \cdot \bec{\nabla}) .$ Hereafter $\Lambda = 0 .$ Thus, in
$ {\bf k} $-space the velocity $ {\bf u} $ is given by
\begin{eqnarray}
u_i({\bf k}) &=& (k / k_{\perp})^{2} [ e_{m} P_{im}({\bf k})
u_{z}({\bf k}) - i ({\bf e} {\bf \times} {\bf k})_{i} \omega({\bf
k}) / k^{2}].
\nonumber \\
\label{C80}
\end{eqnarray}
Multiplying Eq.~(\ref{C80}) for $ u_i({\bf k}_1) $ by $ u_j({\bf
k}_2) $ and averaging over the turbulent velocity field we obtain
\begin{eqnarray}
f_{ij}^{(0)}({\bf k}) &=& (k / k_{\perp})^{4} [f_{\ast}
f^{(0)}({\bf k}) e_{mn} P_{im}({\bf k}) P_{jn}({\bf k})
\nonumber\\
&&  + ({\bf e} {\bf \times} {\bf k})_{i} ({\bf e} {\bf \times}
{\bf k})_{j} G^{(0)}({\bf k}) / k^4] \;, \label{C100}
\end{eqnarray}
where we assumed that the turbulent velocity field in the
background turbulent convection is non-helical. Now we use an
identity
\begin{eqnarray}
(k / k_{\perp})^{2} e_{mn} P_{im}({\bf k}) P_{jn}({\bf k}) =
e_{ij} + k_{ij}^{\perp} - k_{ij}
\nonumber\\
= P_{ij}({\bf k}) - P_{ij}^{\perp}({\bf k}_{\perp}) \;,
\label{C101}
\end{eqnarray}
which can be derived from
\begin{eqnarray*}
k_z (k_z e_{ij} + e_{i} k_{j}^{\perp} + e_{j} k_{i}^{\perp}) =
k_{ij} k^2 - k_{ij}^{\perp} k_{\perp}^{2} \; .
\end{eqnarray*}
Here we also used the identity $ ({\bf k}_{\perp} {\bf \times}
{\bf e})_{i} ({\bf k}_{\perp} {\bf \times} {\bf e})_{j} =
k_{\perp}^{2} P_{ij}^{(\perp)}(k_{\perp}) .$ Substituting
Eq.~(\ref{C101}) into Eq.~(\ref{C100}) we obtain
\begin{eqnarray}
f_{ij}^{(0)}({\bf k}) &=& (k / k_{\perp})^{2} \{f_{\ast}
f^{(0)}({\bf k}) P_{ij}({\bf k}) + [G^{(0)}({\bf k}) / k^2
\nonumber \\
&& - f_{\ast} f^{(0)}({\bf k})] P_{ij}^{\perp}({\bf k}_{\perp}) \}
\; . \label{C102}
\end{eqnarray}
Thus two independent functions determine the correlation function
of the anisotropic turbulent velocity field. In the isotropic
three-dimensional turbulent flow $ G^{(0)}({\bf k}) / k^2 =
f_{\ast} f^{(0)}({\bf k}) $ and the correlation function reads
\begin{eqnarray}
f_{ij}^{(0)}({\bf k}) = f_{\ast} W(k) P_{ij}({\bf k}) / 8 \pi k^2
\; . \label{C103}
\end{eqnarray}
In the isotropic two-dimensional turbulent flow $ G^{(0)}({\bf k})
/ k^2 \gg f_{\ast} f^{(0)}({\bf k}) $ and the correlation function
is given by
\begin{eqnarray}
f_{ij}^{(0)}({\bf k}) = G^{(0)}({\bf k}) P_{ij}^{\perp}({\bf
k}_{\perp}) / k_{\perp}^{2} \; . \label{C104}
\end{eqnarray}
A simplest generalization of these correlation functions is an
assumption that $ G^{(0)}({\bf k}) / [k^2 f_{\ast} f^{(0)}({\bf
k})] - 1 = \varepsilon = const ,$ and thus the correlation
function $ f_{ij}^{(0)}({\bf k}) $ is given by Eq.~(\ref{B8}).
This correlation function can be considered as a combination of
Eqs.~(\ref{C103}) and (\ref{C104}) for the three-dimensional and
two-dimensional turbulence. When $ \varepsilon $ depends on the
wave vector ${\bf k} ,$ the correlation function $
f_{ij}^{(0)}({\bf k}) $ is determined by two spectral functions.

Now we derive Eq.~(\ref{B9}) for the turbulent flux of entropy.
Multiplying Eq.~(\ref{C80}) written for $ u_i({\bf k}_2) $ by $
s({\bf k}_1) $ and averaging over turbulent velocity field we
obtain Eq.~(\ref{B9}). Multiplying Eq.~(\ref{B9}) by $ i ({\bf
k}_{\perp} {\bf \times} {\bf e})_{i} $ we get
\begin{eqnarray}
F^{(0)}({\bf k}) = i ({\bf k}_{\perp} {\bf \times} {\bf e}) \cdot
{\bf \Phi}_{\perp}^{(0)}({\bf k})  \; . \label{C105}
\end{eqnarray}
Now we assume that $ {\bf \Phi}_{\perp}^{(0)}({\bf k}) \propto
{\bf \Phi}^{\ast}_{\perp} f^{(0)}({\bf k}) .$ The integration in $
{\bf k} $-space in Eq.~(\ref{C105}) yields the numerical factor in
Eq.~(\ref{B11}). Note that for simplicity we assumed that the
correlation functions $ F^{(0)}({\bf k}) $ and $ f^{(0)}({\bf k})
$ have the same spectrum. If these functions have different
spectra, it results only in a different value of a numerical
coefficient in Eq.~(\ref{B11}).

Now let us discuss the physical meaning of the parameter   $
\alpha .$ To this end we derived the equation for the two-point
correlation function $ \Phi_{z}^{(0)}({\bf r}) = \langle s({\bf
x}) \, {\bf u}({\bf x}+{\bf r}) \rangle $ of the turbulent flux of
entropy for the background turbulent convection (which corresponds
to Eq.~(\ref{B10}) written in ${\bf k}$-space). Let us we rewrite
Eq.~(\ref{B10}) in the following form:
\begin{eqnarray}
\Phi_{z}^{(0)}({\bf k}) &=& ({\bf \Phi}^{\ast} \cdot {\bf e})
[k^{2} + \Gamma ({\bf e} \cdot {\bf k})^{2}] \tilde \Phi_{w}(k)
\;,
\label{D1}\\
\tilde \Phi_{w}(k) &=& - (3 - \alpha) W(k) / 8 \pi k^{4} \;,
\label{D2}
\end{eqnarray}
where $ \Gamma = 3 (\alpha - 1) / (3 - \alpha) .$ The Fourier
transform of Eq.~(\ref{D1}) reads
\begin{eqnarray}
\Phi_{z}^{(0)}({\bf r}) &=& ({\bf \Phi}^{\ast} \cdot {\bf e})
[\Delta + \Gamma ({\bf e} \cdot \bec{\nabla})^{2}] \Phi_{w}(r) \;,
\label{D3}
\end{eqnarray}
where $ \Phi_{w}(r) $ is the Fourier transform of the function $
\tilde \Phi_{w}(k) .$ Now we use the identity
\begin{eqnarray}
\nabla_{i} \nabla_{j} \Phi_{w}(r) = \psi(r) \, \delta_{ij} + r
\psi'(r) \, r_{ij} \;, \label{D4}
\end{eqnarray}
where $ \psi(r) = r^{-1} \Phi'_{w}(r) $ and $ \psi'(r) = d \psi /
dr .$ Equations (\ref{D3}) and  (\ref{D4}) yield the two-point
correlation function $ \Phi_{z}^{(0)}({\bf r}) :$
\begin{eqnarray}
\Phi_{z}^{(0)}({\bf r}) &=& ({\bf \Phi}^{\ast} \cdot {\bf e})
\biggl(\psi(r)
\nonumber\\
&& + r \psi'(r) {1 + \Gamma \cos^{2} \tilde \theta \over 3 +
\Gamma} \biggr)  \;, \label{D5}
\end{eqnarray}
where $ \tilde \theta $ is the angle between $ {\bf e} $ and $
{\bf r} .$ The function $ \psi(r) $ has the following properties:
$ \psi(r=0) = 1 $ and $ (r \psi')_{r=0} = 0 ,$ e.g., the function
$ \psi(r) = 1 - (r / l_{0})^{q-1} $ satisfies the above
properties, where $ 1 < q < 3 .$ Thus, the two-point correlation
function $ \Phi_{z}^{(0)}({\bf r}) $ of the flux of entropy for
the background turbulent convection is given by
\begin{eqnarray}
\Phi_{z}^{(0)}({\bf r}) &=& ({\bf \Phi}^{\ast} \cdot {\bf e})
\biggl[1 - \biggl({(q - 1) (1 + \Gamma \cos^{2} \tilde \theta)
\over 3 + \Gamma}
\nonumber \\
&&  + 1 \biggr) \biggl({r \over l_{0}} \biggr)^{q-1} \biggr] \; .
\label{D8}
\end{eqnarray}
Simple analysis shows that $ - 3 / (q-1) < \alpha < 3 ,$ where we
took into account that $ \partial \Phi_{z}^{(0)}({\bf r}) /
\partial r < 0 $ for all angles $ \tilde \theta .$ The parameter $
\alpha $ can be presented as $ \alpha = [1 + \xi (q + 1) / (q -
1)] / (1 + \xi / 3) $ and $ \xi = (l_{\perp} / l_{z})^{q-1} - 1 ,$
where $ l_{\perp} $ and $ l_{z} $ are the horizontal $ (\tilde
\theta = \pi / 2) $ and the vertical $ (\tilde \theta = 0) $
scales in which the correlation function $ \Phi_{z}^{(0)}({\bf r})
$ tends to zero. The parameter $ \xi $ describes the degree of
thermal anisotropy. In particular, when $ l_{\perp} = l_{z} $ the
parameter $ \xi = 0 $ and $ \alpha = 1 .$ For $ l_{\perp} \ll
l_{z} $ the parameter $ \xi = - 1 $ and $ \alpha = - 3  / (q-1) .$
The maximum value $ \xi_{\rm max} $ of the parameter $ \xi $ is
given by $ \xi_{\rm max} = q - 1 $ for $ \alpha = 3 .$ Thus, for $
\alpha < 1 $ the thermal structures have the form of column or
thermal jets $ (l_{\perp} < l_{z}) ,$ and $ \alpha > 1 $ there
exist the `'pancake'' thermal structures $ (l_{\perp} > l_{z}) $
in the background turbulent convection.

\section{The identities used for the integration in $ {\bf k} $-space}

To integrate over the angles in $ {\bf k} $-space of
Eqs.~(\ref{A24}) and (\ref{A25}) we used the following identities:
\begin{eqnarray*}
\int k_{ijmn}  \,d \hat \Omega = (4 \pi / 15) \Delta_{ijmn}  \;,
\\
\int k_{ij} \sin^{2} \theta \,d \hat \Omega = (8 \pi / 15) (2
\delta_{ij} - e_{ij}) \;,
\\
\int k_{ijmn} \sin^{2} \theta \,d \hat \Omega = (8 \pi / 105) (3
\Delta_{ijmn} - \Gamma_{ijmn}) \;,
\\
\int k_{ij}^{\perp} \,d \hat \Omega = 2 \pi P_{ij}({\bf e}) \;,
\\
\int k_{ij}^{\perp} k_{mn} \,d \hat \Omega = (\pi / 3)
[P_{ij}({\bf e}) (\delta_{mn} + e_{mn})
\\
+ P_{in}({\bf e}) P_{mj}({\bf e}) + P_{im}({\bf e}) P_{nj}({\bf
e})] \;,
\\
e_{j} \int k_{i}^{\perp} k_{j} k_{mn} k_{\perp}^{-2} \,d \hat
\Omega = (2 \pi / 3) [P_{in}({\bf e}) e_{m} + P_{im}({\bf e})
e_{n}] \;,
\\
e_{j} \int k_{i}^{\perp} k_{j} k_{mn} k^{-2} \,d \hat \Omega = (4
\pi / 3) [P_{in}({\bf e}) e_{m} + P_{im}({\bf e}) e_{n}] \;,
\end{eqnarray*}
where $ P_{ij}({\bf e}) = \delta_{ij} - e_{ij} ,$
$ \, k_{ijmn} = k_{i} k_{j} k_{m} k_{n} / k^{4} ,$
$ \, d \hat \Omega = \sin \theta \,d \theta \,d \varphi ,$
and
\begin{eqnarray*}
\Delta_{ijmn} &=& \delta_{ij}\delta_{mn} +
\delta_{im} \delta_{nj} + \delta_{in}\delta_{mj} \;,
\\
\Gamma_{ijmn} &=& \delta_{ij} e_{mn} + \delta_{im} e_{jn} + \delta_{in} e_{jm}
+ \delta_{jm} e_{in}
\\
&& + \delta_{jn} e_{im} + \delta_{mn} e_{ij} \;,
\\
\gamma_{ijm} &=& \Delta_{ijmn} e_{n} = \delta_{ij} e_{m} + \delta_{im}
e_{j} + \delta_{jm} e_{i}  \;,
\\
e_{m} \gamma_{ijm} &=& \delta_{ij} + 2 e_{ij} \;,
\\
e_{n} \Gamma_{ijmn} &=& \gamma_{ijm} + 3 e_{ijm}  \;, \quad
e_{mn} \Gamma_{ijmn} = \delta_{ij} + 5 e_{ij} \;,
\\
P_{ij}({\bf k}) &+& \varepsilon P_{ij}^{\perp}({\bf k}_{\perp})
= (1 + \varepsilon) \delta_{ij} - \varepsilon e_{ij} - k_{ij}
- \varepsilon k_{ij}^{\perp} \; .
\end{eqnarray*}
The above identities allow us to calculate the following integrals
(which were used for the derivation of equations for the
Reynolds stresses and turbulent flux of entropy):
\begin{eqnarray*}
\int \tau(k) k_{ij} \Phi_{m}^{(0)}({\bf k}) \, d {\bf k} = ({\bf
\Phi}^{\ast} \cdot {\bf e}) (\tau_{0} / 30) [15 \delta_{ij} e_{m}
\\
+ 10 \alpha \, e_{ijm} - (2 \alpha + 3) \gamma_{ijm} + 6 b_{ijm}]
\;,
\\
\int \tau(k) k_{mn} f_{ij}^{(0)}({\bf k}) \, d {\bf k} = (f_{\ast}
\tau_{0} / 6) [(\varepsilon / 4) (\Gamma_{ijmn} - e_{ijmn})
\\
- (1/5 + \varepsilon / 4) \Delta_{ijmn} + (1 + \varepsilon)
\delta_{ij} \delta_{mn}
\\
- (\varepsilon / 2) \delta_{ij} e_{mn} - \varepsilon e_{ij}
\delta_{mn}] \;,
\end{eqnarray*}
where
\begin{eqnarray*}
b_{ijm} =  \delta_{ij} (\Phi^{\ast}_{\perp})_{m} +
[\varepsilon_{iml} \, ({\bf \Phi}^{\ast} {\bf \times} {\bf e})_{j}
+ \varepsilon_{jml} \, ({\bf \Phi}^{\ast} {\bf \times} {\bf
e})_{i}] e_{l} \;,
\end{eqnarray*}
and we used an identity $ e_{q} \varepsilon_{pqn} ({\bf
\Phi}_{\ast} {\bf \times} {\bf e})_{m} \Delta_{ijmn} = b_{ijp} .$

\end{document}